\documentclass[review,twoside,12pt]{elsarticle}

\usepackage{lineno,hyperref}

\journal{Progress in Particle and Nuclear Physics}

\usepackage{epsfig}

\usepackage[utf8]{inputenc}
\usepackage[english]{babel}
\usepackage{amsmath}
\usepackage{amsfonts}
\usepackage{amssymb}
\usepackage{graphicx}	
\usepackage{dcolumn}	
\usepackage{bm}			
\usepackage{hyperref}	
\usepackage{todonotes}
\usepackage{upgreek}
\usepackage[english]{babel}
\usepackage{bm}
\usepackage{color,soul}
\usepackage{threeparttable}
\usepackage[margin=1in]{geometry}

\begin{document}

\begin{frontmatter}

\title{Recent progress in laser spectroscopy of the actinides}

\author[mymainaddress,mysecondaryaddress,mytertiaryaddress]{Michael Block\corref{mycorrespondingauthor}}
\cortext[mycorrespondingauthor]{Corresponding author}
\ead{m.block@gsi.de}

\author[mytertiaryaddress]{Mustapha Laatiaoui}

\author[mymainaddress,mysecondaryaddress]{Sebastian Raeder}

\address[mymainaddress]{GSI Helmholtzzentrum f\"ur Schwerionenforschung, Planckstrasse 1, 64291 Darmstadt, Germany}
\address[mysecondaryaddress]{Helmholtz-Institut Mainz, Staudingerweg 18, 55118 Mainz}
\address[mytertiaryaddress]{Johannes Gutenberg-Universit\"at Mainz, Fritz-Strassmann-Weg 2, 55128 Mainz, Germany}

\begin{abstract}
The interest to perform laser spectroscopy in the heaviest elements arises from the strong impact of relativistic effects, electron correlations and quantum electrodynamics on their atomic structure. Once this atomic structure is well understood, laser spectroscopy also provides access to nuclear properties such as spins, mean square charge radii and electromagnetic moments in a nuclear-model independent way. This is of particular interest for the heaviest actinides around $N=152$, a region of shell-stabilized deformed nuclei. The experimental progress of laser spectroscopy in this region benefitted from continuous methodological and technical developments such as the introduction of buffer-gas-stopping techniques that enabled the access to ever more exotic nuclei far-off stability. The key challenges faced in this endeavor are small yields, nuclides with rather short half-lives and the need to search for atomic transitions in a wide spectral range guided by theoretical predictions. This paper describes the basics of the most common experimental methods and discusses selected recent results on the atomic and nuclear properties of the actinides up to nobelium where pioneering experiments were performed at the GSI Helmholtzzentrum f\"ur Schwerionenforschung in Darmstadt, Germany.
\end{abstract}


\end{frontmatter}


\section{Introduction}

In recent years the investigation of radioactive nuclides by laser spectroscopy has been booming. The subject has been covered in several review articles, for example by Kluge and N{\"o}rtersh{\"a}user \cite{Kluge2003}, by Cheal and Flanagan \cite{Cheal2010}, by Campbell, Moore, and Pearson \cite{Campbell2016} and by N{\"o}rtersh{\"a}user \cite{Noertershaeuser2019}.  A review on the perspectives for laser spectroscopy of the heaviest elements was given by Backe {\it{et al.}} \cite{Backe2015} including off-line measurements on Fm. A more general overview addressing the determination of ground-state properties of  radioactive isotopes by atomic-physics techniques was given by Blaum, Dilling and N{\"o}rtersh{\"a}user \cite{Blaum2013}.

Laser spectroscopy of radionuclides has been used to investigate the nuclear structure evolution in exotic nuclei and contributed to the understanding of new nuclear structure features that have been found in exotic nuclei with extreme proton-to-neutron ratio in the past. A prominent example in light nuclei is $^{11}$Li whose size is about the same as for the spherical nucleus $^{208}$Pb with many more nucleons \cite{Tanihata_1985}. With laser spectroscopy one was able to shed light on its detailed structure \cite{Sanchez2006}. 
Another well-known example of a nuclear structure feature that has even been discovered with the help of laser spectroscopy is the nuclear shape staggering in neutron-deficient Hg isotopes (around $N=104$), which was revealed in experiments at ISOLDE for the first time \cite{Bonn1972} and which was only recently extended \cite{Marsh2018}. 
Meanwhile studies of shape coexistence form an entire subfield of nuclear physics, see for example the review by Heyde and Wood \cite{Heyde2011} and references therein.

The progress in laser spectroscopy towards more exotic nuclides was possible thanks to the development of new techniques with highest sensitivity, the continuous improvement of the efficiency of established methods, an increased availability of laser systems in a wide spectral range and the availability of ever more radionuclides far-off stability. The latter is due to the upgrades of existing and the upcoming of next-generation radioactive ion beam (RIB) facilities. 

The regions in the nuclear chart in which laser spectroscopy has been performed are governed by the production scheme. Initially, laser spectroscopy works were performed at facilities where a wide range of nuclei are produced and delivered as mass-separated low-energy beams. This concerns isotope separator on-line (ISOL) facilities with hot-cavity ion source like ISOLDE (ISOL Device) at CERN, Geneva, Switzerland \cite{Kugler2000,Borge2017,Neugart2017,Fedosseev2017}, ISAC (Isotope Separator and Accelerator) at TRIUMF (Tri-University Meson Facility) at Vancouver, Canada \cite{Bricault2008,Dilling2014,Mane2011,Voss2013}, and the former Holifield Radioactive Ion Beam Facility (HRIBF) at Oak Ridge National Laboratory (ORNL) in Oak Ridge, TN, USA \cite{Beene2011,Miernik2013,Liu2016}. Many development works for in-source laser spectroscopy have also been performed at the Petersburg Nuclear Physics Institute (PNPI) \cite{Panteleev2005}.

However, the ISOL method has some drawbacks. As it relies on radionuclides effusing out of an ion source, it has restrictions with respect to the half-life of nuclides that can be studied, and it prevents experiments on refractory elements. Therefore, methods such as the Ion-Guide Isotope Separation On-Line (IGISOL) technique at the department of physics' accelerator laboratory (JYFL) in Jyv{\"a}skyl{\"a}, Finland \cite{Forest_2012,Moore_2013,Moore2013,Vormawah2018} were developed. This technique ensures a fast extraction of ions that are stopped in an inert gas and thus gives access to shorter-lived nuclides and  allows furthermore extending laser spectroscopy to refractory elements. A similar approach was used at the former Leuven Isotope Separator On-Line (LISOL) facility at Louvain-la-Neuve, Belgium \cite{Kudryavtsev2013,Cocolios2009,Ac_ferrer2017towards}.

In order to extend the reach towards even more radionuclides laser spectroscopy techniques have also been implemented at fast-beam fragmentation facilities, for example at the Institute of Physical and Chemical Research (RIKEN) in Wako, Japan \cite{Nakamura2006}, and at the National Superconducting Cyclotron Laboratory (NSCL) in East Lansing, MI, USA with the Beam Cooler for Laser Spectroscopy (BECOLA) setup \cite{Minamisono2013}. Also at the future FAIR facility in Darmstadt, Germany, laser spectroscopy is planned within the LaSpec project of the NUSTAR collaboration \cite{Rodriguez2010,Herlert2017}. 

At the KEK Isotope Separation System (KISS) in Japan laser spectroscopy is performed utilizing multi-nuclear transfer reactions to produce neutron-rich nuclides around $N=126$ \cite{Hirayama2017,Hirayama2020,Choi_2020}. Laser spectroscopy on ions produced by the fission of a $^{252}$Cf source inside a gas cell of the  CAlifornium Rare Isotope Breeder Upgrade (CARIBU) facility at Argonne National Laboratory, USA is also being prepared \cite{Maas2017}.

Despite an increasing number of facilities and a large suite of experimental approaches the access to actinides is limited.
The employed nuclear reactions in the ISOL approach can be used to produce a few actinide isotopes of Ac, Np and Pu, for example. On the contrary, long-lived isotopes up to Fm can be produced in macroscopic quantities on the picogram to milligram scale in nuclear reactors by neutron-capture reactions followed by $\beta^{-}$ decay.  This facilitates off-line measurements often with conventional laser spectroscopy approaches after laborious radiochemical separation and purification.

The heavier actinides as well as the transactinides have to be produced on-line at accelerator facilities in nuclear fusion-evaporation reactions. This way of production typically results in low yields of often only single atoms at a time. Moreover, the production is limited to a number of neutron-deficient actinide isotopes by the available projectile-target combinations. In addition, the reaction products are delivered at kinetic energies of tens of megaelectronvolt.

A crucial ingredient to enable laser spectroscopy under these conditions was the introduction of adequate beam-preparation techniques involving buffer-gas stopping and advancing the IGISOL approach. The introduction of such techniques that enabled the access to exotic nuclides of essentially all elements by exploiting almost all types of nuclear reactions, including fast-beam fragmentation and fusion evaporation reactions. With these new techniques low-energy beams of exotic nuclides suitable for laser spectroscopy can be prepared with beam characteristics comparable to atomic beams previously available only at ISOL facilities. 
Based on these recent developments the region of the actinide elements has been studied more intensely in recent years. Pioneering experiments have for example been performed on short-lived Am fission isomers by Backe {\it{et al.}} \cite{Backe1998} and on $^{255}$Fm by Sewtz {\it{et al.}} \cite{Sewtz2003}. Select other recent experiments up to the heaviest element studied by laser spectroscopy so far, nobelium, are discussed later in this article.

The scientific motivation for laser spectroscopy of the actinides is driven by the interest in their atomic structure, which is strongly influenced by relativistic effects, quantum electrodynamics and electron correlations. These effects often lead to several close-lying levels and, sometimes, to a change of the ground state configuration, not only in comparison to their lighter homologues but also within the series \cite{Eliav2015,Pershina2015,Schwerdtfeger2015}. 
The (first) ionization potential (IP) is one of the key observables that reflects these effects and the change in chemical properties.

At the same time the actinides' nuclear structure that is affected by the competition of the Coulomb repulsion and the stabilization by nuclear shell effects draws attention \cite{Giuliani2019}. The actinide isotopes known to date span the region from the proton drip line, the well-established spherical shell closure at $N=126$ up to the region around the weak deformed shell closure at $N=152$. Therefore, most of the actinide nuclei are well deformed and exhibit interesting features, among others, octupole deformation and the occurrence of $K$ isomers, which could be longer lived than their ground-state counterparts.

In very heavy and superheavy nuclei the strong Coulomb repulsion within the nucleus is predicted by nuclear theory calculations to cause a central depression in the proton distribution of several percent that may eventually result in so-called bubble nuclei \cite{Schuetrumpf2017}.

Signatures of the shape and size of nuclei, in the ground state or an isomeric state, are reflected in the charge distribution of the nucleus probed by laser-driven electronic transitions. Thus, laser spectroscopy may also provide information on the nuclear shell structure. The key properties are the changes in the mean-square nuclear charge radii, which can be deduced from the isotope shift of atomic energy levels and often show kinks around shell closures, and the nuclear spins and moments that can be inferred from the hyperfine splitting. The comparison of measured electromagnetic moments to theoretical predictions, for example by the nuclear shell-model, provides rich information on the underlying structure. Experimental access to these observables is a challenging task for nuclei far-off stability that are often short-lived and only available in smallest amounts.

Even though information on the deformation can be obtained by gamma spectroscopy, for example, by the observation of rotational bands, the interpretation of the data requires input from nuclear models. In contrast, laser spectroscopy can provide information in a nuclear-model independent way. Other established methods such as electron scattering and muonic X-rays remain not applicable to exotic radionuclides. Some efforts to extend these methods to radioactive nuclei are being made, for example in the context of the Self-Confining Radioactive Ion Target (SCRIT) project for electron scattering at the RIKEN RIB factory \cite{Ohnishi2015}, for muonic X-rays at the RIKEN-RAL Muon Facility \cite{Strasser2001}, and within the muX project at the Paul Scherrer Institut, Switzerland \cite{Skawran2019}.

\begin{figure}
    \centering
        \includegraphics[width=1.0\textwidth]{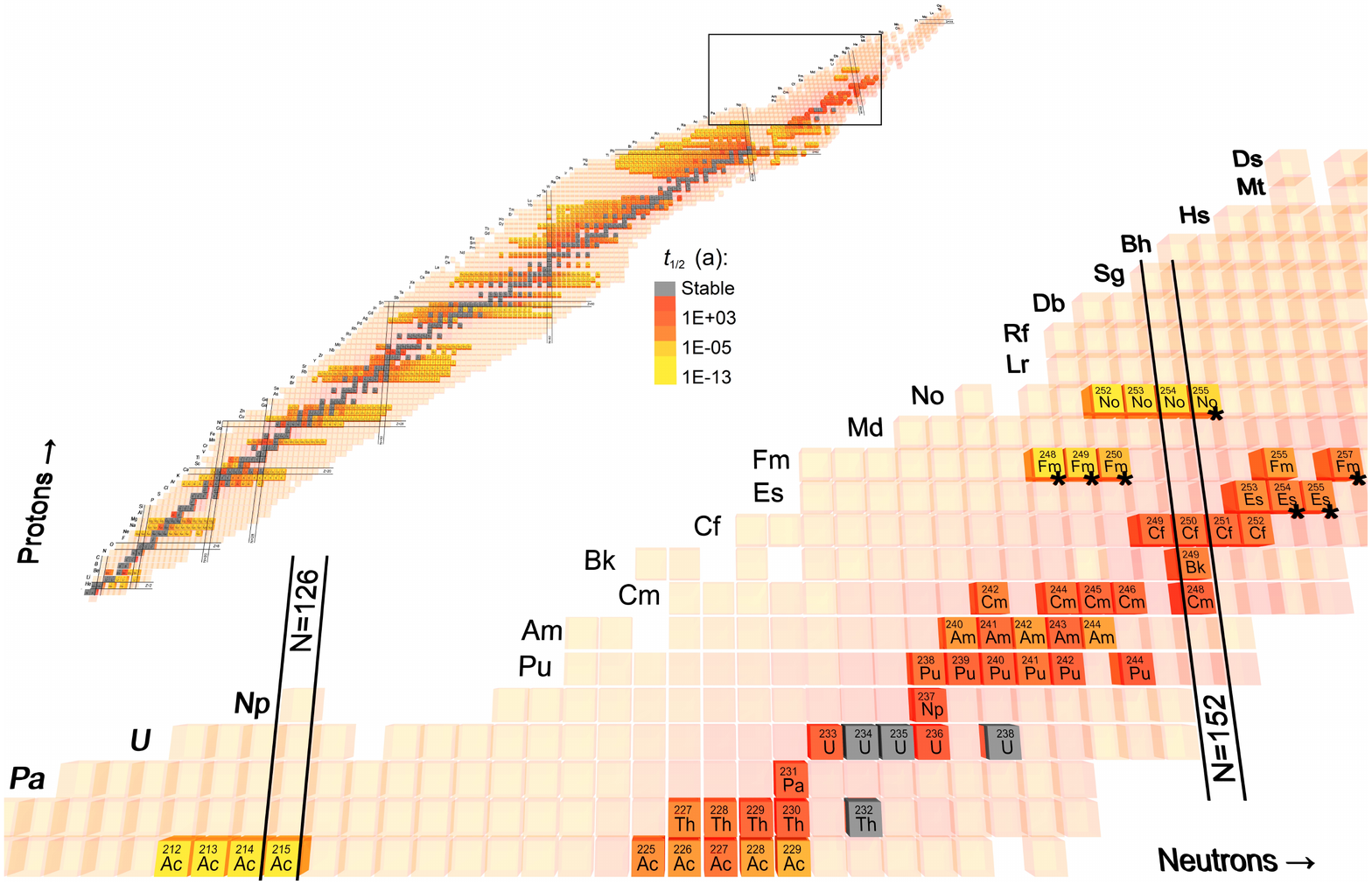}
    \label{fig:chart}
    \caption{Nuclear Chart in the region of the heaviest elements showing those isotopes which have been addressed by laser spectroscopy so far (color-highlighted boxes) and their half-lives (color coded). Measured but not yet reported nuclides are marked with stars.}
\end{figure}

In the case of lacking reference isotopes, some nuclear properties can still be extracted from laser spectroscopy data if reliable atomic calculations for hyperfine parameters and other atomic properties are available. Here a particular challenge arises in the heavy actinide and in the transactinide region, where relativistic and quantum electrodynamic effects as well as electron correlations strongly influence the electronic structure \cite{Eliav2015}, but at the same time no stable isotopes are available as references.

This article summarizes the recent progress in the field of laser spectroscopy of the (heaviest) actinides. The region of interest and the work performed in recent years is summarized in figure~\ref{fig:chart}.
The manuscript is organized as follows. In section~\ref{sec:atomic} the atomic structure aspects are addressed. In section~\ref{sec:nuclear} the determination of nuclear properties from the laser spectroscopy data is discussed. In section~\ref{sec:basics} the basics of laser spectroscopy are presented. Sections~\ref{sec:buffer_gas}-\ref{sec:gas-jet} discuss specifically buffer-gas-stopping and related laser spectroscopy techniques. In section~\ref{sec:production} the different production schemes for the actinides are summarized. In section~\ref{sec:results} we present selected recent results illustrating the benefit of the various methods for measurements of atomic and nuclear properties starting from actinium reaching up to nobelium isotopes investigated at the GSI in section~\ref{sec:nobelium}. Future perspectives arising from novel approaches are addressed in section~\ref{sec:perspectives}.

\section{Atomic structure of the heaviest elements}
\label{sec:atomic}

The electronic structure of the atom is unique for each element and is responsible for fundamental atomic and chemical properties. Probing interatomic transitions by laser spectroscopy helps to elucidate this structure, allowing one to determine atomic level energies, transition strengths and lifetimes in addition to basic properties such as ionization potentials.

Glen T. Seaborg came up with the actinide concept in the 1940s at a time when many elements beyond uranium were synthesized for the first time. His concept was based on the chemical behavior of the actinide elements \cite{Seaborg1945,Seaborg1946,Seaborg2012} and the similarity to their chemical homologues, the lanthanides. In the lanthanides the 4f orbital is filled with up to 14 electrons. The actinide series starts with actinium (Ac, $Z=89$), and terminates with lawrencium (Lr, $Z=103$). Relevant atomic orbitals for the ground state configurations are 5f, 6d, 7s and 7p, which give access to a large number of electron configurations and thus to a large number of atomic levels. Within the series, the ground state configuration often changes from an element to another as the 5f, 6d, and 7p orbitals are rather close in energy due to the above-mentioned relativistic effects.

In general, the atomic structure of most actinide elements is only partially known due to the scarcity of isolated material and the lack of abundant isotopes with suitable half-lives.
\begin{figure}[t]
    \includegraphics[width=1.0\textwidth]{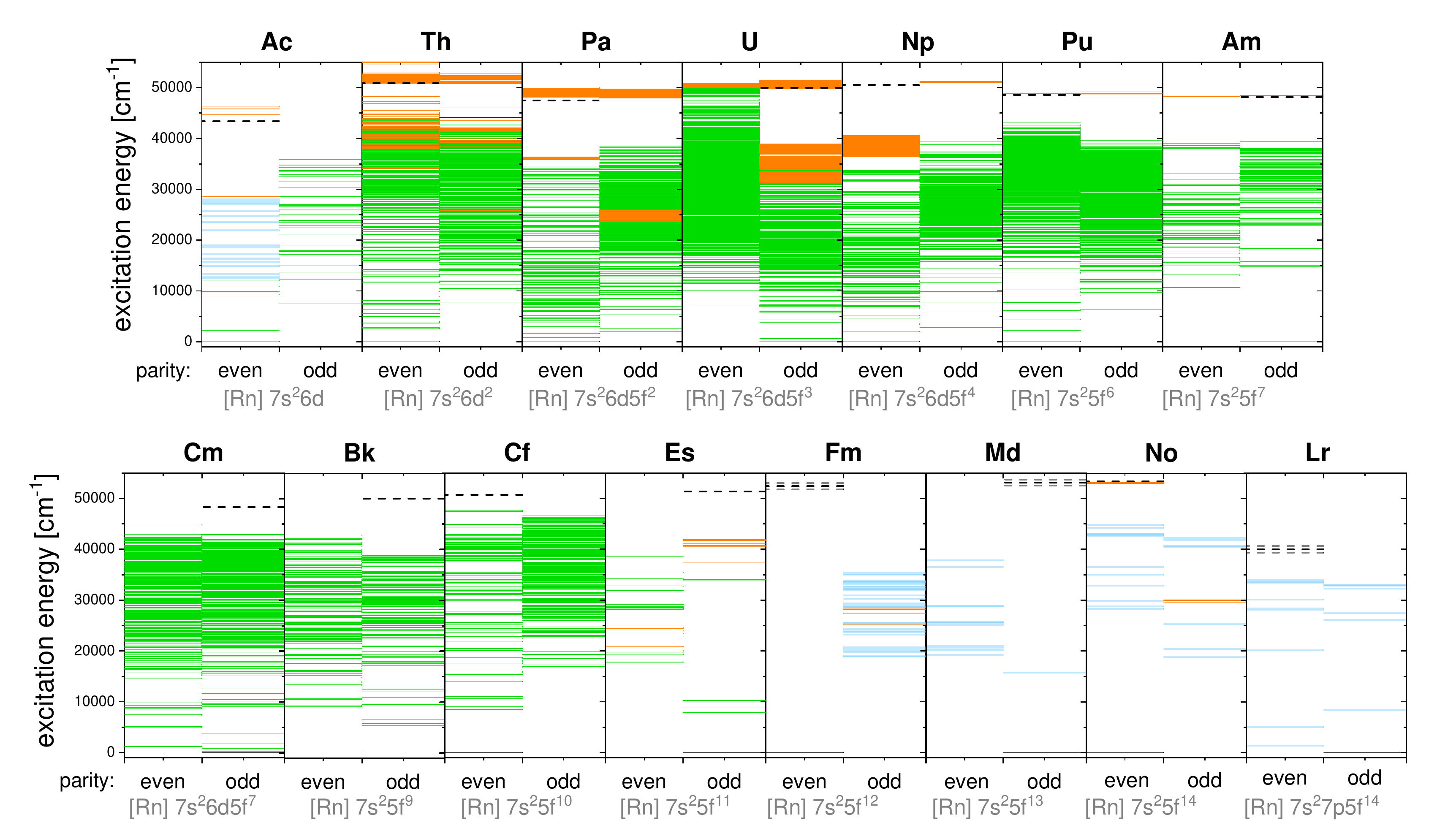}
    \caption{\label{fig:levels} Overview on atomic levels of neutral actinide species ordered by parity. The data is taken from the compilation by Blaise and Wyart\cite{blaise1992selected} (green).  More recently, additional levels (orange) were reported for Ac \cite{Ac_waldek2000bestimmung,Ac_Zhang_2020}, Th \cite{Th_Johnson1992,Th_Redman2014,Th_raeder2010a}, Pa \cite{Naubereit2018}, U \cite{U_Bajaj1988,U_Broglia1983,U_Mago1987,U_Mago1987a,U_Mago1988,U_Mago1988a,U_Manohar1989,U_Miyabe2000,U_Miyabe2002,U_Ray1990,U_Ray1992,U_Rodrigues2000,U_Shi2000,U_Smyth1991,U_Suri1987,U_bushaw2007a,Raeder2010spurenanalyse}, Pu \cite{Ac_waldek2000bestimmung,Raeder2010spurenanalyse,Am_Kneip2020}, Am \cite{Am_Kneip2020,Raeder2019b}, Es \cite{Es_Wyart2005}, Fm \cite{Sewtz2003,Sewtz2003b,Backe2005}  and No \cite{Laatiaoui2016,Chhetri2018}. In addition, theoretical predictions are shown (blue) for Ac \cite{Dzuba2019a}, Fm \cite{Allehabi2019,Es_Wyart2005}, Md \cite{Li2020}, No \cite{Borschevsky2007a,Liu2007,Dzuba2014,Dzuba_2019,Indelicato2007,Fritzsche2005,Beerwerth2019} and Lr \cite{Borschevsky2007b,Fritzsche2007,Dzuba2014}  }
\end{figure}
Figure\,\ref{fig:levels} shows a summary of atomic levels for the actinides from actinium to lawrencium about which only a little is known \cite{Sato2015}.

While energy levels in the region up to 40\,000\,cm$^{-1}$ are reported, the knowledge of higher-lying levels, which are important for resonant laser ionization as discussed later, is scarce.
From einsteinium (Es, $Z$=99) onwards, the amount of information is quickly diminishing, a result of the reduced availability of these elements from breeding in nuclear reactors.
An extended overview of the atomic structure with an emphasis on optical and laser spectroscopic techniques is given in \cite{Worden2010}.

The transactinides start with the element rutherfordium (Rf, $Z$=104) at which the 5f-shell is closed and the 6d shell is successively filled up to copernicium (Cn, $Z=112$).
According to the classical arrangement reflected in the periodic table of elements the p-shell is filled from nihonium (Nh, $Z$=113) till oganesson, (Og, $Z$=118), to date the heaviest element known, which should exhibit a completely filled 7p shell as by analogy with noble gases.
Nevertheless, direct relativistic effects of orbital contraction of (especially) s- and p$_{1/2}$ orbitals close to the nucleus and likewise the indirect effects of energy destabilization of d- and f-orbitals due to better shielding of the nuclear Coulomb potential, are  expected to alter the chemical behavior significantly compared to classical expectations \cite{Schwerdtfeger2015}.
For example, the elements copernicium and flerovium (Fl, $Z$=114) were predicted to behave like noble gases \cite{Pitzer1975}.
In view of the limited experimental data, Cn seems to follow the chemical trend within its group, for Fl some deviations, however, were observed \cite{Yakushev2014}.
The gas-phase chemistry of the transactinides has been reviewed recently \cite{Tuerler2015}.

For the understanding of the atomic structure of elements around Fm, atomic calculations help in guiding experimental investigations, while for other heavier elements that are presently inaccessible for optical spectroscopy, theoretical investigations are currently the only way to obtain information on their fundamental properties.
Experimental exploration of the atomic structure and characterization of atomic levels in turn can test the predictive power of theoretical model calculations and thereby may validate model assumptions. 
Accurate theoretical calculations require a fully relativistic description and are challenged by electron-electron correlations, which increase the computational effort tremendously, in particular, for systems with open shells.
Different frameworks for theoretical investigations include Multi-Configuration Dirac-Hartree Fock (MCDHF) calculations \cite{Joensson2007,Grant2007}, Fock-space coupled cluster (FSCC) \cite{Eliav2015} and different variations of the Configuration Interaction (CI) method \cite{Safronova2009,Dzuba2014a}.
The theory for the atomic structure of the heaviest elements has been reviewed by Eliav {\it{et al.}} \cite{Eliav2015}. Meanwhile, the relative uncertainties from these calculations can reach a 1\%-level for excitation energies and ionization potentials.
Calculations of atomic transitions of heavy elements including isotope shifts were performed in an astrophysical context to enable the search for atomic transitions in star light to determine the elemental abundances \cite{Dzuba2017}.
Progress in the theoretical treatment of open shells enables now calculation of elements with partly filled 5f shells such as Fm \cite{Allehabi2019}, Md \cite{Li2020} and for elements with open d-shells \cite{Lackenby2019,Lackenby2020}.
At present, calculations for most elements are existing \cite{Dzuba2016,Dzuba2016b,Dzuba2016d,Dzuba2019a,Dzuba_2019,Geddes2018,Lackenby2018a,Kahl:2019,Dinh2016,Eliav2019}
with some special emphasis on Og \cite{Jerabek2018,Lackenby2018}, the heaviest element known to date, which concludes the 7th row of the periodic table of chemical elements.

\subsection{Deriving nuclear properties from laser spectroscopy data}
\label{sec:nuclear}

A key interest in laser spectroscopy of radionuclides in general and of the actinides in particular is to obtain information on the nuclear structure evolution far-off stability. This has been covered in several reviews \cite{Kluge2003,Cheal2010,Campbell2016} and is also subject of a recent review by Flanagan {\it{et al.}} \cite{Flanagan2020}. The idea is that the electronic energy levels and their shifts and splittings carry information on the charge distribution of the nucleus. The splitting of a transition due to hyperfine interaction carries the information of the nuclear spin and the nuclear dipole and quadrupole moments. As mentioned earlier, this information can be obtained from laser spectroscopy in a nuclear model-independent way.

If the hyperfine splitting
\begin{equation}
\label{eqn:hfs}
\Delta E_\textrm{\small HFS}=A\frac{C}{2}+B\frac{(3/4)C(C+1)-I(I+1)J(J+1)}{2I(2I-1)J(2J-1)}
\end{equation}
\begin{equation}
\textrm{with}~~
C=F(F+1)-I(I+1)-J(J+1),~~
A=\mu \frac{B_e(0)}{IJ},~~\textrm{and}~~
B=eQ_\textrm{s} \left\langle\frac{\partial^2 V}{\partial z^2}\right\rangle_0~
\end{equation}
can be resolved then the splitting pattern and the individual Racah intensities of the hyperfine peaks allow one to determine the nuclear spin $I$. Additional complications may arise from optical pumping effects.

The hyperfine structure (HFS) results from the coupling of the electron angular momentum $J$ with the nuclear spin $I$ resulting in levels with a total angular momentum $F$ where the size of the splitting is determined by the hyperfine parameters $A$ and $B$.
These hyperfine parameters are linked to the interaction of the magnetic field induced by the electrons, $B_e(0)$, with the nuclear magnetic moment $\mu $ and to the electric-field gradient, $\left\langle\frac{\delta^2 V}{\delta z^2}\right\rangle_0$, interaction with the spectroscopic quadrupole moment $Q_\textrm{s}$ of the nucleus.
To obtain the nuclear moments, the electronic parts of these interactions have to be known. These latter can be inferred either from HFS measurements on reference isotopes with known nuclear moments or otherwise from atomic calculations.

The isotope shift of an optical transition is linked to the interaction of the atomic orbitals with the effective nuclear volume. Measuring this shift across an isotopic chain enables one to probe changes in the nuclear size and shape as reflected in the nuclear charge distribution.
For example, the isotope shift $\delta \nu ^{A,A'}$ $= \nu ^{A'} - \nu ^{A}$ of an atomic transition in different isotopes $A$ and $A'$ of a given element provides a measure of the change in the mean square charge radius $\delta \langle r^2 \rangle^{A,A'}$ according to
\begin{equation}
\label{Eq_radii}
\delta \nu ^{A,A'} = \frac{m^{A'}-m^A}{m^{A'} m^A} M + F_s \left( \delta \langle r^2 \rangle^{A,A'} + \frac{C_2}{C_1} \delta \langle r^4 \rangle^{A,A'} + ... \right) ,
\end{equation}
where $M$ and $F_s$ are the mass and the field-shift factors, respectively, and $C_i$ are the expansion parameters to be determined by theory~\cite{Seltzer1969}.
While the mass dependence in front of the mass-shift factor $M$ gets very small for heavy elements \cite{Noertershaeuser2019}, the determination of the field shift is large since it increases roughly proportional to $Z^2$.
In the assessment of the field shift contribution, often one can neglect the higher order terms of the charge radii changes by considering a constant electron wave function across the nuclear volume. In Eq.~\ref{Eq_radii}, however, we give a more general form of the field shift~\cite{Seltzer1969} as such an approximation becomes significantly less valid for the heaviest elements.
A fully experimental calibration of the field shift factor $F_s$ and the mass-shift factor $M$ requires at least three reference isotopes with known changes in their mean square charge radii.

In order to obtain absolute radii at least one reference value for a specific isotope is needed. Such data can be obtained for example from electron scattering. However, this is usually only applicable to stable, or at least extremely long-lived nuclides. 
In general, reference data is rather rare for elements above plutonium. 
Only recently there have been groups envisaging muon or electron scattering, for example on $^{248}$Cm \cite{Skawran2019}.

Thus, atomic theory becomes essential, this time, for the extraction of nuclear properties, in particular for the very heavy elements.
The influence of the nuclear shape can be modeled in atomic theory and allows the prediction of the electronic factors \cite{Cheal2012}.
The present accuracy for the extraction of nuclear properties from these atomic calculations can reach a few percent at best. The uncertainty can be estimated from a comparison to atomic systems with a similar valence electron configuration and, additionally for MCDHF, the convergence for an increasing configuration space can be evaluated \cite{Beerwerth2019}, while for the configuration-interaction technique a perturbative treatment of an increased configuration space is available \cite{Dzuba2017b}.

Starting from the neutron shell closure at $N=126$, nuclei of the heavy elements have an increasingly deformed shape until the deformed shell closure at $N=152$ \cite{Heenen2015}. In addition, they are predicted to show a central depression of the charge distribution, which contributes to the effective charge radius \cite{Schuetrumpf2017}.
In this context, some discussion on a proper consideration of these effects was raised \cite{Dzuba2017,Flambaum2018,Flambaum2019,Allehabi2020}.
Accurate theoretical calculations are therefore necessary to improve the understanding of the nuclear structure of the heaviest elements.

\section{Laser spectroscopy}
\subsection{Principles of resonant laser excitation}
\label{sec:basics}

\begin{figure}[t]
    \centering
 \includegraphics[width=0.8\textwidth]{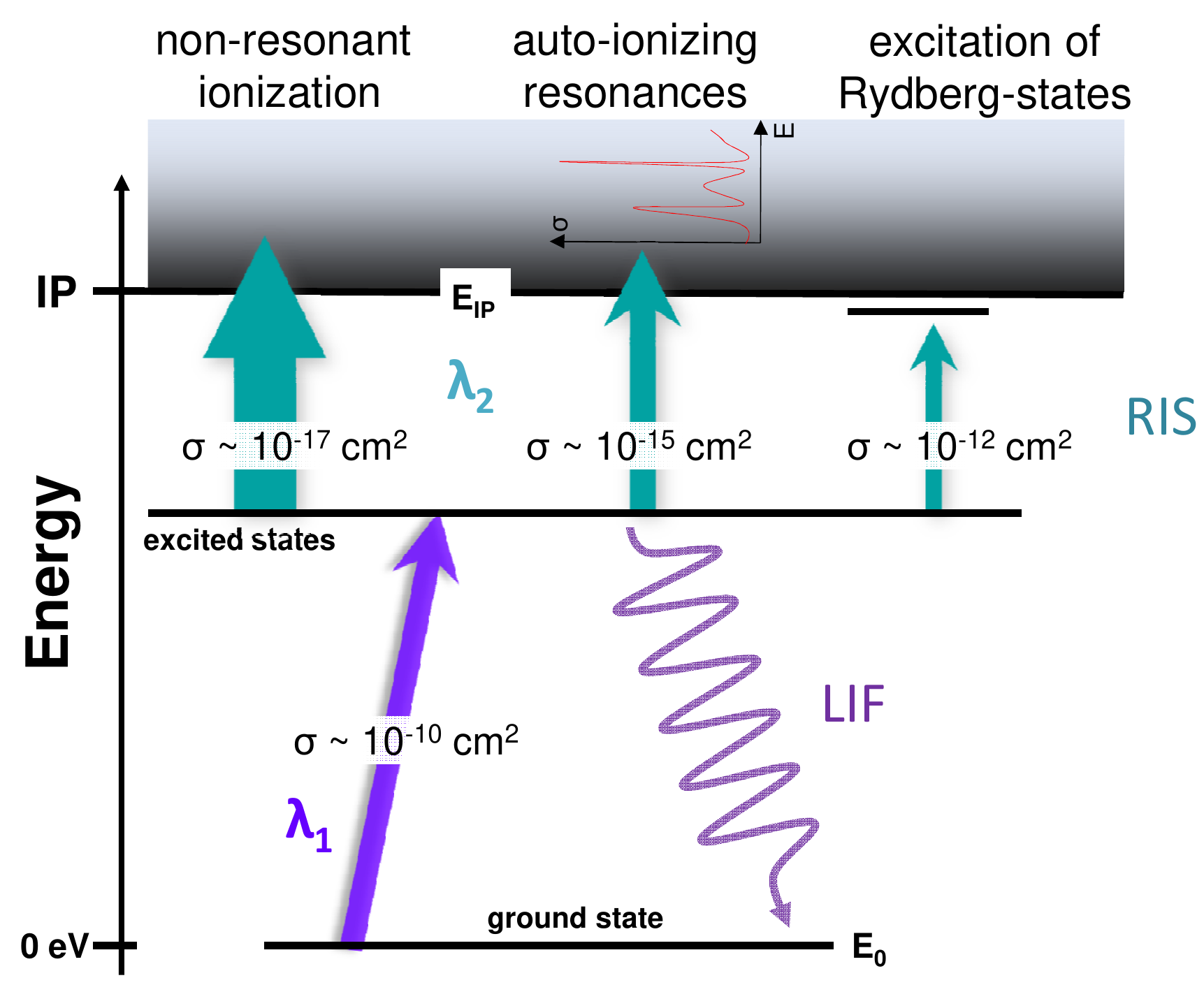} \caption{\label{fig:ion_scheme}Different two-step ionization schemes and corresponding  cross sections, figure adapted from \cite{Raeder2020}. Following a first resonant laser excitation ($\lambda_1$) to an excited state the second excitation step ($\lambda_2$) can be performed either non-resonant but with high laser power (left), by utilizing auto-ionizing resonances (middle) or by excitation to Rydberg states close to the continuum (right). In the latter case the remaining energy necessary for ionization may be provided for example by collisions or field ionization.}
\end{figure}

In laser spectroscopy resonant laser excitations are exploited to measure properties of atomic transitions such as excitation energies, lifetimes and hyperfine structure splittings, thereby revealing the electronic structure of an element.
Following a laser excitation, there are two main principles for detecting optical resonances: one by detecting fluorescence from radiative decays of the excited atoms, and a second by using a sequence of laser excitations to detach one electron from the atom and finally to detect the charge-state change of the produced photo-ions.
These two principles are schematically depicted in figure\,\ref{fig:ion_scheme}.
Spectroscopy using laser-induced fluorescence (LIF) is a widely used technique for many applications \cite{Demtroeder2014}, e.g., for laser spectroscopy of laser-ablated uranium \cite{Harilal2020}.

The combination of a collinear arrangement of laser light with an accelerated beam enables laser spectroscopy of exotic nuclei with high precision \cite{Campbell2016,Anton1978,Neugart2017}.
The fast beam reduces the relative velocity spread of the ions extracted from an ion source and thus the Doppler broadening down to a few tens of MHz, which for strong transitions may correspond to the natural linewidth.
The sensitivity is limited by the detection efficiency of the emitted photons and by the background rate of the photon detection.
The latter can be significantly reduced when using bunched ion beams in conjunction with gated photon detection \cite{Nieminen2002}.

Resonance ionization spectroscopy (RIS) is a technique explored by Lethokov and Hurst \cite{letokhov1987laser,Hurst1988a}. A valence electron is stepwise resonantly excited beyond the continuum such that a different charge state results. One of the simplest RIS methods uses a two-step two-color photoionization of usually neutral atoms as schematically presented in figure~\ref{fig:ion_scheme}. The method features a high efficiency due to the large cross sections for resonant excitation processes and a high sensitivity due to efficient ion manipulation and detection. The method also provides an inherent element selectivity.
These advantages make the RIS technique attractive for many applications from trace detection \cite{lu2003laser,hurst1988b,Franzmann2017,Raeder2019b} to efficient ionization of rare isotopes at radioactive beam facilities  \cite{Alkhazov1989,Vermeeren1994,Fedosseev2012,Marsh2014a} and for spectroscopic investigations \cite{Alkhazov1992,Campbell2016}.

One of the limiting factors for the method's efficiency is the final ionizing step.
Compared to the efficiency of resonant excitations with a cross section $\sigma_{\textrm{res}}=\frac{\lambda^2}{2\pi}\approx 10^{-10}\,\textrm{cm}^{2}$ the cross section for non-resonant laser ionization (the final excitation into the continuum above the ionization potential) is several orders of magnitude smaller ($\sigma_{\textrm{cont}}\approx 10^{-17}\,\textrm{cm}^{2}$).
Sometimes, alternative scenarios for the final ionization step are utilized like resonant excitation via high-lying bound Rydberg states ($\sigma_{\textrm{Ry}}\approx 10^{-12}\,\textrm{cm}^{2}$) or auto-ionizing (AI) states ($\sigma_{\textrm{AI}}\approx 10^{-15}\,\textrm{cm}^{2}$) above the IP. However, these latter scenarios are often not practical for initial level searches in the actinides with barely known excitation schemes.

A benefit from involving Rydberg states is that they allow one to determine the IP precisely as demonstrated for many elements including some actinides, like recent measurements on At, Ac and No \cite{Rothe2013a,Rossnagel2012,Chhetri2018}. A drawback of Rydberg states is their sensitivity to ambient conditions such as gas density and electric fields.

AI resonances are short-lived bound states above the IP involving more than one electron, which then decay, leaving behind an ion--electron pair. They may result in enhanced resonant ionization, making them an ideal tool for RIS applications.
However, not all elements feature AI resonances and, in general, the knowledge on these structures is limited also theoretically. 

\subsubsection{Laser ionization in a hot cavity and in an atomic beam}

\begin{figure}[t]
 \centering
 \includegraphics[width=0.9\textwidth]{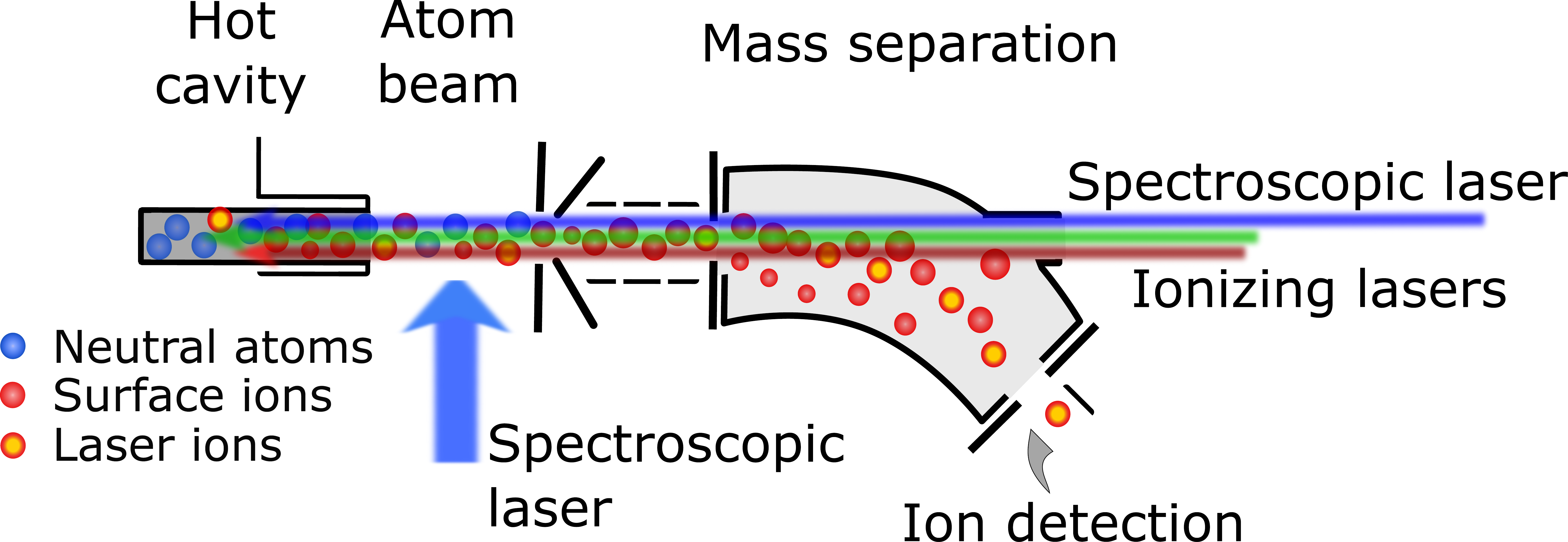} 
 \caption{\label{fig:hotcavity} Schematic setup of RIS inside a hot cavity when ionizing laser beams are applied in a counter-propagating direction along the hot cavity axis and inside the emerging atomic beam with perpendicularly aligned laser radiation.}
\end{figure}

At ISOL facilities the exotic nuclei from the target are released into a hot surface ion source cavity.
These hot cavities are made of a refractory metal with high melting point. For elements with a low first ionization potential a high surface ionization efficiency and high ion yields can be achieved \cite{Zandberg1959}. The ions are usually accelerated to 30--60 keV kinetic energy and then mass separated using electromagnetic separators. The mass resolving power ranges from m/$\Delta$m = 300 to about 10,000. This is often insufficient to resolve nuclear isobars.
Injecting resonant laser radiation into this hot furnace as shown in Fig.~\ref{fig:hotcavity} allows one to resonantly excite and subsequently photo-ionize elements with a suitably larger ionization potential \cite{kluge1985laser,Koester2003}.
Due to the confinement of the evaporated neutrals in the cavity the interaction time with the laser radiation is prolonged resulting in an increased ionization efficiency.
Depending on the length of the furnace and the thermal velocities of the released atoms a laser system with a repetition rate of $10\,$kHz is well suited for efficient ionization.
Besides the mere selective ionization of exotic isotopes this geometry can be used for laser spectroscopy, where the resulting ion yield is monitored as a function of the applied laser frequencies.

This technique is not limited to on-line application but can also be employed off-line with stable and long-lived samples \cite{Liu2020,Rossnagel2012}.
The resolution is limited by the optical Doppler effect from the thermal velocities of the atoms in the cavity.
For an actinide with mass $238\,$u in a cavity heated to $2000\,$K the resulting Doppler width for an atomic transition at $370\,$nm is about $2\,$GHz (FWHM), which matches well the bandwidth of typical pulsed laser radiation, enabling a realization of ion sources for on-line facilities and for off-line trace analysis applications \cite{Marsh2014a,Raeder2012}.

The resolution can be improved by employing the laser spectroscopy not inside the hot cavity but in the atomic beam emerging from the furnace or a heated filament at the cost of a reduced total efficiency. A part of the efficiency loss can be regained when employing the ionization in an radiofrequency guiding field \cite{Heinke2016}.

The ISOL Facility at the Japan Atomic Energy Agency (JAEA) in Tokai, Japan \cite{Osa2008}, is presently the only one of its kind that employs fusion-evaporation reactions for the production of radioisotopes of actinides, introduces them into an ISOL-type ion source by a gas jet system and then utilizes surface ionization to produce actinide beams up to lawrencium \cite{Sato2015}. In the past similar production schemes were used at other facilities such as the GSI on-line mass separator \cite{Roeckl2012} and the LISOL facility \cite{Huyse1992}, in the latter case also for the production of Pa and Th isotopes.

\subsubsection{Collinear laser spectroscopy techniques}

In this approach a fast ion beam can be neutralized by charge exchange, for example with an alkali vapor, and is then (anti-)collinearly overlapped with the laser beam(s). Typically either the fluorescence light is observed \cite{Campbell2016,Anton1978} or RIS ions are detected \cite{Schulz1991,Cocolios2013}.
Due to the fast atomic beam the residual Doppler broadening is effectively suppressed while the optical Doppler shift may be used to scan the resonance by tuning the ion beam energy instead of scanning the laser frequency.
This enables reaching a high spectral resolution (close to the natural linewidth) for optical transitions across long chains of isotopes, e.g. at ISOL facilities \cite{Neugart2017}.

This high resolution is often beneficial for the determination of nuclear parameters from the obtained spectra, but hampers a wide-range search for atomic levels. Limitations of the method arise from its reduced sensitivity in the case of photon detection, and the requirement of a high-quality, low-energy ion beam.

As presently actinide elements are preferably produced in other methods than ISOL production processes that do neither provide high-quality low-energy beams nor sufficient yield, the application of collinear laser spectroscopy methods for actinide elements requires additional preparation to obtain a suitable ion beam using, e.g. buffer-gas cells or hot-cavity ion sources, while the low production yield challenges the sensitivity of the methods.

One example{for the potential of collinear laser spectroscopy on actinides is given by the recent off-line studies where an ion beam of long-lived Pu isotopes was obtained from evaporation and resonant ionization of reactor-produced material from a filament in a buffer-gas cell at JYFL with the aim to probe the change in nuclear charge radii \cite{Pu_voss2017}.
Collinear laser spectroscopy after preparation of ion beams from buffer-gas ion catchers has recently also been implemented at in-flight facilities \cite{Minamisono2013} but has not been employed to actinides yet.

\subsection{Laser spectroscopy utilizing buffer-gas cells}
\label{sec:buffer_gas}

In this section methods for laser spectroscopy utilizing a buffer-gas cell are briefly discussed. A detailed discussion on this topic relevant for actinides and transactinides has been previously addressed by Backe {\it{et al.}} \cite{Backe2015}. 

One of the requirements for most laser spectroscopy methods that are applied to radionuclides is their preparation as beams with small energy spreads of a few eV and with total kinetic energy on the order of $\leq 60$~keV. In particular for the production schemes in which the radionuclides emerge at energies of tens of MeV the reaction products have to be first slowed down by utilizing so-called buffer-gas stopping cells or ion catchers \cite{Kudryavtsev1996,Dendooven1997,Wada2003,Weissman2004,Neumayr2006a,Morrissey2007,Savard2008,Savard2011,Plass2013,Kaleja2020,Schwarz2020,Sumithrarachchi2020}. 

In these devices, radionuclides are thermalized in an inert buffer gas, usually helium or argon, at gas pressures of about 30-500\,mbar depending on the kinetic energies of the incoming ions. Prior to entering the buffer-gas environment, the ions lose most of their kinetic energy in a solid degrader, which is often also utilized as an entrance window separating the buffer-gas environment from the vacuum section of the preceding transport beamline. Depending on the ionization potential of the ion of interest compared to the ionization potential of the buffer-gas atoms a significant fraction of the slowed down radionuclides remain charged, either as singly, doubly or sometimes triply charged ions \cite{Schury2017,Wense2016b}. A crucial prerequisite for this is a high purity of the buffer gas with impurities on a ppb level. Thus, the employed device should fulfill ultrahigh vacuum requirements and use purification methods to assure highest cleanliness.

A first laser spectroscopy experiment on Be ions from fragmentation that were slowed down in a buffer-gas cell was performed at RIKEN \cite{Nakamura2006}.
Besides application for on-line produced nuclides also off-line studies were performed, e.g., on plutonium isotopes  \cite{Pohjalainen2016}. Here filaments with long-lived plutonium isotopes were prepared in Mainz, similar to the filaments used for the investigation of the ionization potential of actinides \cite{Erdmann:1998}. They were used for evaporation of Pu atoms in a gas cell at JYFL in Jyv{\"a}skyl{\"a} where laser ionization was applied. The resulting beam of plutonium ions was used for collinear laser spectroscopy \cite{Pu_voss2017}.

\subsection{RADRIS - Radioactive Decay Detected Resonance Ionization Spectroscopy}
\label{sec:radris}

\begin{figure}[t]
 \centering
 \includegraphics[width=0.6\textwidth]{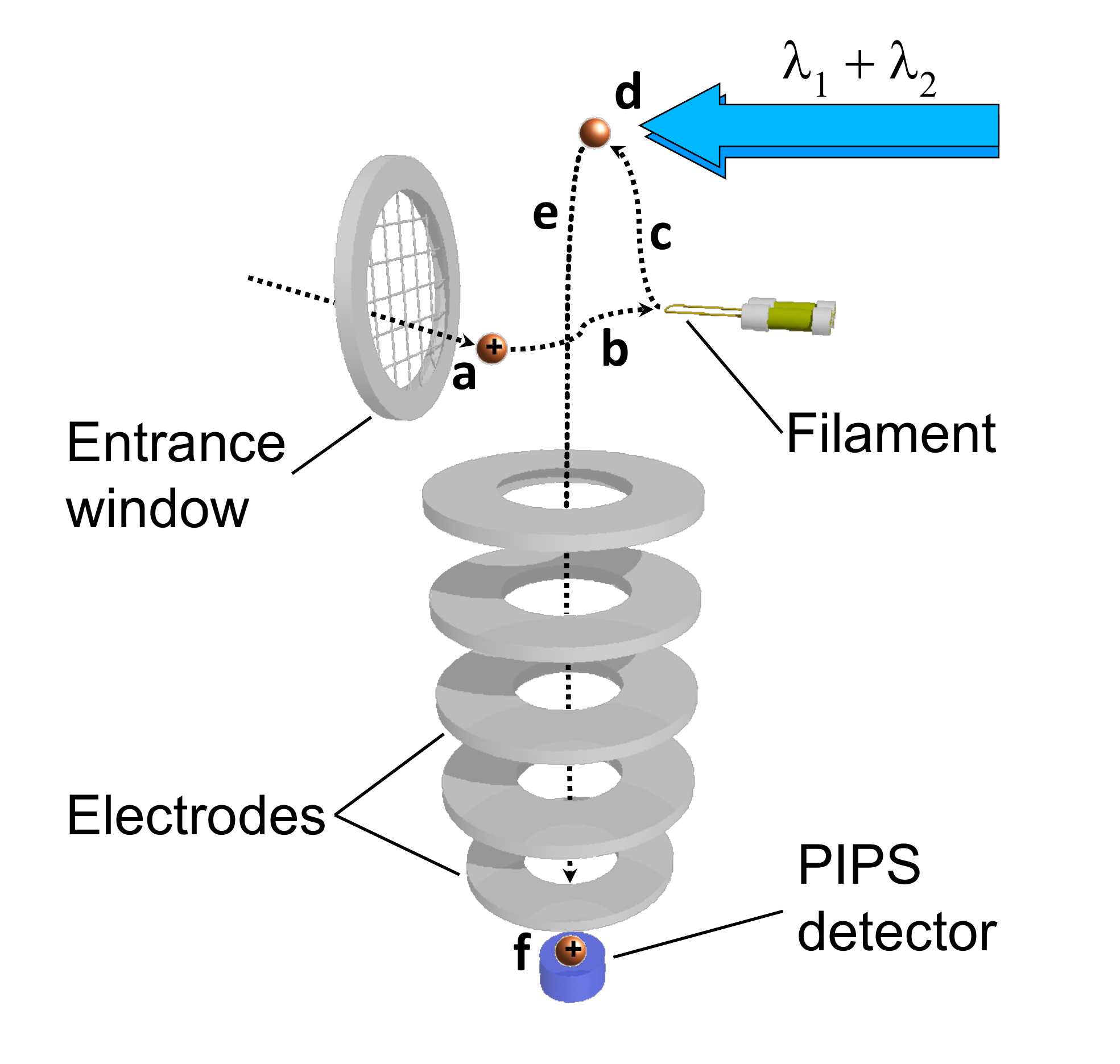}
 \caption{\label{fig:RADRIS}Principle of the RADRIS technique. a, Thermalization of fusion products in buffer gas; b, accumulation on the filament; c, re-evaporation from the filament; d, two-step photoionization of neutral atoms using two laser beams of different wavelengths $\lambda_1$ and $\lambda_2$; e, accumulation of re-ionized fusion products on the Passivated, Implanted, Planar Silicon (PIPS) detector; f, radioactive-decay detection.}
\end{figure}

In contrast to the use of buffer-gas cells as sources for thermalized ions, often neutral atoms are required for laser spectroscopy, in particular in RIS-based experiments. A crucial step is thus the efficient neutralization of the nuclides of interest. As discussed above the major part of the slowed down particles remains charged in a high-purity inert gas. Thus, a neutralization scheme has to be employed prior to the resonant laser ionization. 

One option is using a filament of a suitable material on which the ions can be collected with the help of electric fields. This method, the Radiation Detected RIS method (RADRIS), has been suggested for the element nobelium and demonstrated for ytterbium by Backe {\it{et al.}} \cite{Backe2007}. In this technique the nuclides of interest are neutralized on a tantalum filament and then evaporated from the filament by a short heating pulse. The evaporation as neutral atoms is element dependent and requires a careful selection of the filament material. Ideally, the material should tolerate elevated temperatures and at the same time should not promote surface ionization. 

In the RADRIS method \cite{Backe2007,Lautenschlaeger2016} at GSI, the radionuclides of interest are slowed down in argon gas and accumulated by electrostatic fields on the tip of a filament wire, see Fig.~\ref{fig:RADRIS}. After pulse heating the filament, the neutral atom cloud from the filament expands in the gas and is photo-ionized in a two-step laser excitation. 

The relatively slow diffusion in the buffer gas confines the atom cloud spatially and results in a good spatial overlap with the laser beams. The photo-ions created during this process are guided to a silicon detector for radioactive-decay detection.
Thus, the method features a high efficiency, but the spectral resolution is limited by gas collisions that may also result in collision-induced quenching.
 
A high sensitivity is achieved due to suppressed background from non-neutralized particles and because the detection of laser ions is based on detecting the characteristic radioactive decay of the nuclides of interest. This puts some constraints on the accessible half-life range. The method works well for alpha-decaying nuclides with half-lives of about 1--1000~s.

Depending on the decay mode a (significant) fraction of the decay daughters remains on the filament and becomes accessible for laser spectroscopy as well. Decay modes with low recoil energy such as electron capture decay are particularly suited. This opens up access to nuclides that are difficult or impossible to produce directly. Proof-of-principle experiments have been performed at the GSI. For example, the electron capture decay of $^{255}$Lr has been exploited to produce $^{255}$No and measure its hyperfine structure with RADRIS.

\subsection{Laser ionization in a gas jet}
\label{sec:gas-jet}

\begin{figure}[t]
 \centering
 \includegraphics[width=0.9\textwidth]{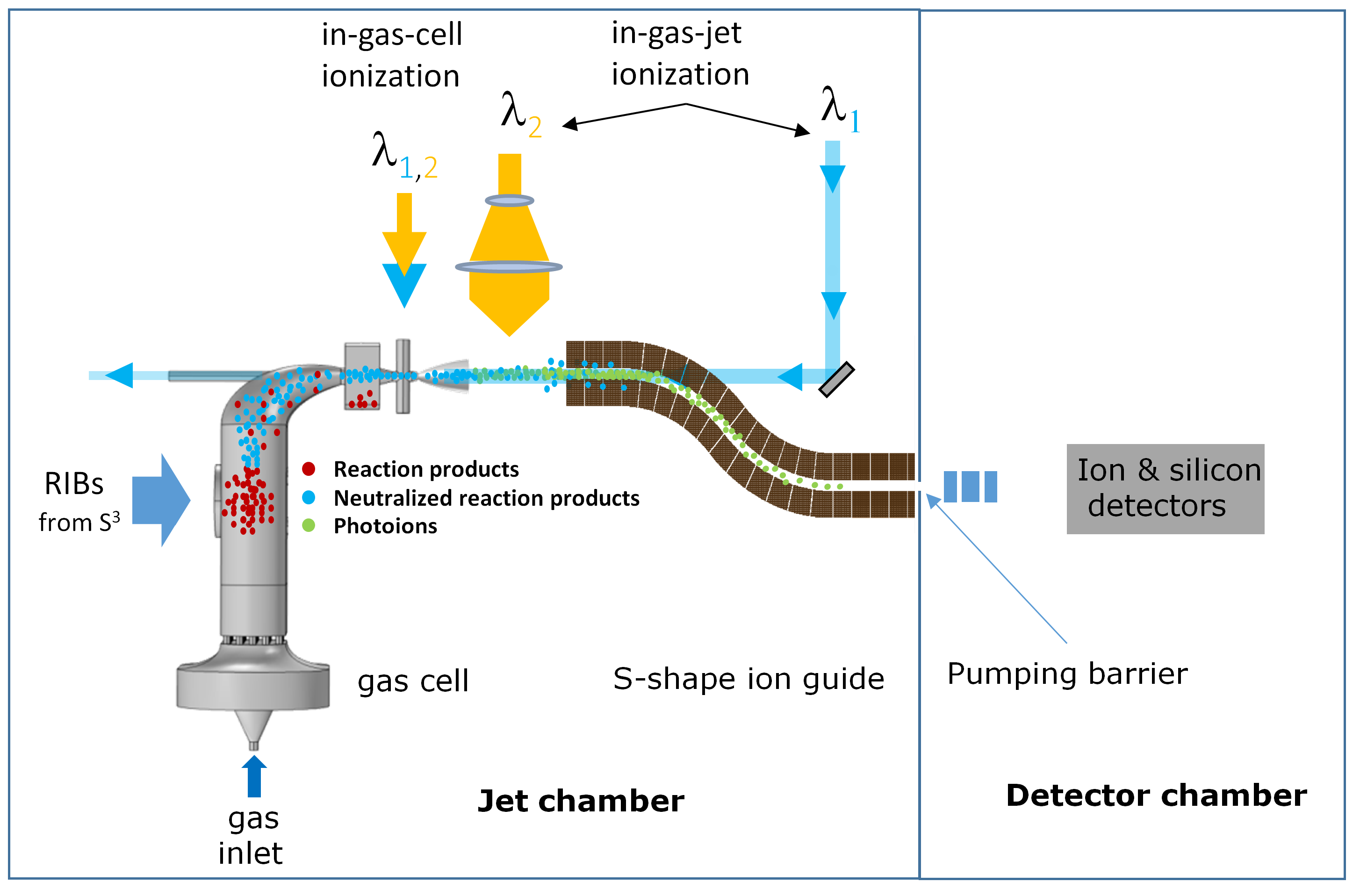}
 \caption{\label{fig:gas_jet}Schematic of the gas-jet setup intended for the low-energy branch at the Super Separator Spectrometer S$^3$ at GANIL.  After separation by S$^3$ the recoils are stopped in a gas cell and transported by the gas flow to the exit with a de Laval nozzle. The atoms in the formed jet are illuminated by the laser light and the photo ions are extracted and transported by a radiofrequency quadrupole with S-shape to allow for a collinear alignment of the laser beam with the gas jet. The figure was provided by the courtesy of R. Ferrer.}
\end{figure}

Another technique in the context of heavy elements is the laser spectroscopy in a well-collimated, supersonic gas jet emerging from a high pressure gas stopping cell \cite{Kudryavtsev2013}.
The usage of a supersonic gas jet to improve the spectral resolution for laser spectroscopy on atoms and molecules is an established technique, see, e.g., \cite{Smalley1977}. However, the usage of a well-formed gas jet in combination with a high-repetition-rate laser system for laser spectroscopy is a new development, which enables achieving a high spectral resolution while maintaining the necessary high efficiency.

A sketch of the setup planned for the future Super Separator Spectrometer S$^3$ at the Grand Acc\'el\'erateur National d’Ions Lourds (GANIL), France is shown in figure\,\ref{fig:gas_jet}. In this technique recoiling ions of heavy elements are separated in flight after production in a fusion-evaporation reaction. After passing a thin entrance window, they are subsequently stopped, thermalized and neutralized in argon gas at a pressure of about $300\,$mbar. The more plasma-like conditions prevail during stopping the more neutralization becomes efficient. Alternative neutralization schemes may use radioactive $\beta$ sources providing electrons~\cite{Ferrer2013}. The atoms are guided by the gas flow to the exit of the gas stopping cell, where a de Laval nozzle forms a well-collimated supersonic gas jet.

Due to the low density and temperature in such a gas jet, the resolution for laser spectroscopy improves considerably.
The velocity of the atoms is mainly determined by the extraction velocity of the argon carrier gas to about $500\,$m/s. Therefore, a region for the interaction of the laser light with the jet of a length of $55\,$mm in combination with a $10\,$kHz repetition rate laser system may enable addressing all atoms in the jet. The produced laser ions are then captured in a radiofrequency ion guide and transported to the detection region.

A proof-of-principle study was carried out at the Leuven ISOL separator. It demonstrated that high-resolution spectroscopy with a high efficiency on neutron deficient Ac isotopes produced on-line is indeed feasible \cite{Ac_ferrer2017towards}.
At KU Leuven, Belgium, a new laboratory was recently established to drive the development of the gas jet technique \cite{Kudryavtsev2016}. This new technique enabled a detailed characterization of gas jets created by different nozzle geometries \cite{Zadvornaya2018}.
Dedicated apparatuses are developed to employ laser spectroscopy in a supersonic gas jet at the in-flight separator S$^3$ at GANIL \cite{Ferrer2013,Raeder2020}, the Mass Analysing Recoil Apparatus (MARA) separator at Jyv{\"a}skyl{\"a} University \cite{Papadakis2018,Papadakis2016} and in a modified version using ion-guiding electric fields and neutralization on filaments at the Separator for Heavy-Ion reaction Products (SHIP) of the GSI \cite{Raeder2020a}.

\section{Production of actinide isotopes}
\label{sec:production}

Albeit all actinide isotopes are radioactive, the isotopes $^{232}$Th, $^{235}$U and $^{238}$U are primordial and are still present in macroscopic amounts on earth due to their long half-lives.
Furthermore, a very low abundance of primordial $^{244}$Pu was reported \cite{HOFFMAN1971}. Trace amounts of the isotopes in primordial radioactive decay chains exist naturally such as $^{227}$Ac, $^{228}$Ac, $^{227}$Th, $^{228}$Th, $^{231}$Pa, $^{234}$Pa, $^{234}$U as well as isotopes produced in naturally occurring nuclear reactions such as $^{236}$U \cite{Steier2008}.
All other radionuclides, especially of the heavier elements, have to be produced artificially in nuclear reactions.

\subsection{Spallation and fragmentation}

A bombardment of a heavy target with high-energy protons is usually used to break up target nuclei in competing processes of spallation, fragmentation and fission. This in combination with a large distribution of nucleon evaporation of the highly excited nuclei results in a variety of nuclides that can be produced for a given target \cite{Duppen2006}.
As the production process is based on shattering target nuclei the produced unstable nuclides are lighter and therefore only uranium and thorium targets allow for the production of actinide isotopes with $Z \leq 92$ and $N \leq 146$.

ISOL facilities use a thick target producing significant amounts of exotic nuclei.
To release the produced nuclei from the target matrix, the target container is heated to high temperatures. The heating and release processes are sensitive to physical properties of the sample material and of the target container.
The nuclei then diffuse into an ion source region where different ionization techniques such as surface ionization, electron impact ionization and laser ionization can be employed, see Sec.~\ref{sec:basics}.
Two major conventional ISOL facilities that are presently operational are ISOLDE \cite{Borge2017} and TRIUMF \cite{Dilling2014}. 
Future facilities are presently under construction, for example SPES (Selective Production of Exotic Species) at Legnaro, Italy \cite{Prete2012} and RAON (Rare isotope Accelerator complex for ON-line experiment) at Daejeon, Korea \cite{Kang2013}.

\subsection{Production of long-lived actinides in nuclear reactors}
\label{sec:reactor}

The light actinide isotopes can be produced in nuclear reactors by exploiting neutron capture reactions followed by $\beta^-$ decay. Based on the neutron capture cross section, the half-lives, and the decay modes (branchings) a certain set of isotopes is available. A reactor cycle of up to several months can be chosen depending on the half-lives of the targeted actinide isotopes to be produced. In this way one can produce very long-lived isotopes of heavier actinides eventually up to Fm. This production scheme stops at this element due to the lack of $\beta^-$ decay channels in presently known Fm isotopes.

Some of the accessible long-lived actinide isotopes of Np, Pu, Am, Cm, Bk, Cf, Es and Fm are produced in picogram to milligram quantities at high-power nuclear reactors with a thermal neutron flux on the order of $2 \times10^{15}$ neutrons / (cm$^2$ s). Such reactors comprise the High Flux Isotope Reactor (HIFR) reactor at ORNL \cite{Roberto2015} and the reactor at the Research Institute for Advanced Reactors (RIAR) at Dmitrovgrad, Russia \cite{Karelin1997}. Both are utilized to produce actinide samples for research and industry. Usually the reactors are operated to produce commercial isotopes by neutron irradiation of mixed americium and curium targets in campaigns.

One of the actinide isotopes that is of commercial interest is $^{252}$Cf.
The production process also delivers other actinide isotopes as byproducts, which can be used for fundamental research. After irradiation in the reactor and a suitable cool down period a radio-chemical separation is performed.
A selection of the actinide isotopes that are available in rather large quantities is summarized in table~\ref{tab:act-isotope} using the ORNL as an example.

\begin{table}[htp]
 \caption{Long-lived actinide isotopes that are available at  ORNL, TN, USA; adapted from \cite{Roberto2015}.}
 \label{tab:act-isotope}
 \begin{center}
  \begin{tabular}{|c|c|c|c|}
   \hline
   Nuclide & Half-life / years & Inventory / mg & Isotopic Enrichment \\
   \hline
   $^{237}$Np & $2.144\times10^6$  & 1000 & $>99\%$  \\
   $^{242}$Pu & $3.75 \times10^5$  & 5500 & $>99\%$  \\
   $^{241}$Am & 432.6  & 3500 &$>99\%$ \\
   $^{243}$Am & 7364  & 1000 & $>99\%$  \\
   $^{244}$Cm & 18.1  & 1000 & $>90\%$ \\
   $^{248}$Cm & $3.48\times10^5$  & 2500 & 80-95\% \\
   $^{249}$Bk & 330 days & - & $>99\%$ \\
   $^{249}$Cf & 351  & 170 & $>99\%$  \\
   $^{251}$Cf & 898  & 150 & $\approx35$\% \\
   \hline
  \end{tabular}
 \end{center}
\end{table}%

One of the difficult isotopes to produce in this context is $^{249}$Bk, which can be obtained in 10--20~mg quantities despite its half-life of only about 330 days. Like many other long-lived isotopes from breeding processes also this one has been used as target material in the synthesis of superheavy elements \cite{Roberto2015,Oganessian2010,Oganessian2012,Khuyagbaatar2014}.  

Also, long-lived actinide isotopes are of great interest for laser spectroscopic experiments. 
Isotopes that are available in microgram quantities or more can be studied using conventional techniques. More challenging are those that are available only in picogram to nanogram amounts such as $^{253-255}$Es and $^{255,257}$Fm. 
$^{255}$Fm, for instance, became available in the past for off-line experiments at the University of Mainz, Germany. Despite its half-life of only about 20 hours, Sewtz {\it{et al.}} were able to identify seven atomic transitions in this element for the first time after scanning a large spectral range based on predictions from atomic theory \cite{Sewtz2003}. The longer-lived Fm isotope $^{257}$Fm with a half-life of about 100 days was also produced at ORNL and recently studied in Mainz. 
Besides sample quantities the radiochemical purity is another critical factor for laser spectroscopic studies. Elevated radioactivity from unwanted byproducts often limits sample handling in laser laboratories in terms of the sample size.

\subsection{Production of actinides in fusion-evaporation reactions}

Many actinide isotopes that are neither produced in the ISOL approach nor can be bred in nuclear reactors can be accessed by fusion-evaporation reactions at accelerator facilities. The elements beyond Fm and likewise the superheavy elements can presently only be produced in this way. To this end a projectile is accelerated to energies close to the Coulomb barrier that is typically in the range of 4--6 $A$MeV and then impinges on a thin target of about 0.5 mg/cm$^2$ area density to enable a nuclear fusion process with the target nuclei to take place. In the early days, light-ion induced reactions with protons, deuterons or alpha particles were used to synthesize new elements up to Cf \cite{Fry2013} soon after the first cyclotron became available at Lawrence Berkeley National Laboratory (LBNL) in Berkeley, USA. Later-on heavy-ion induced reactions were exploited to create heavier elements. The elements Es and Fm were initially discovered in the debris of nuclear explosions \cite{Ghiorso1955,Meierfrankenfeld2011}. Short-lived neutron-deficient Es and Fm isotopes can nowadays also be produced by fusion-evaporation reactions.

Nuclear fusion reactions are often distinguished based on the excitation energy of the compound nucleus as 'hot' and 'cold' fusion reactions. In cold-fusion reactions with beams from Ne to Zn reacting with Pb and Bi targets neutron-deficient isotopes of the elements uranium up to nihonium, $Z=113$, may be produced. The name cold fusion was given referring to the excitation energy of the compound nucleus of typically less than 20~MeV resulting in the evaporation of one or two neutrons. Several new elements up to Nh were discovered in such reactions \cite{Hofmann2000,Morita2015a}. 

\begin{figure}[t]
 \centering
  \includegraphics[width=0.8\textwidth]{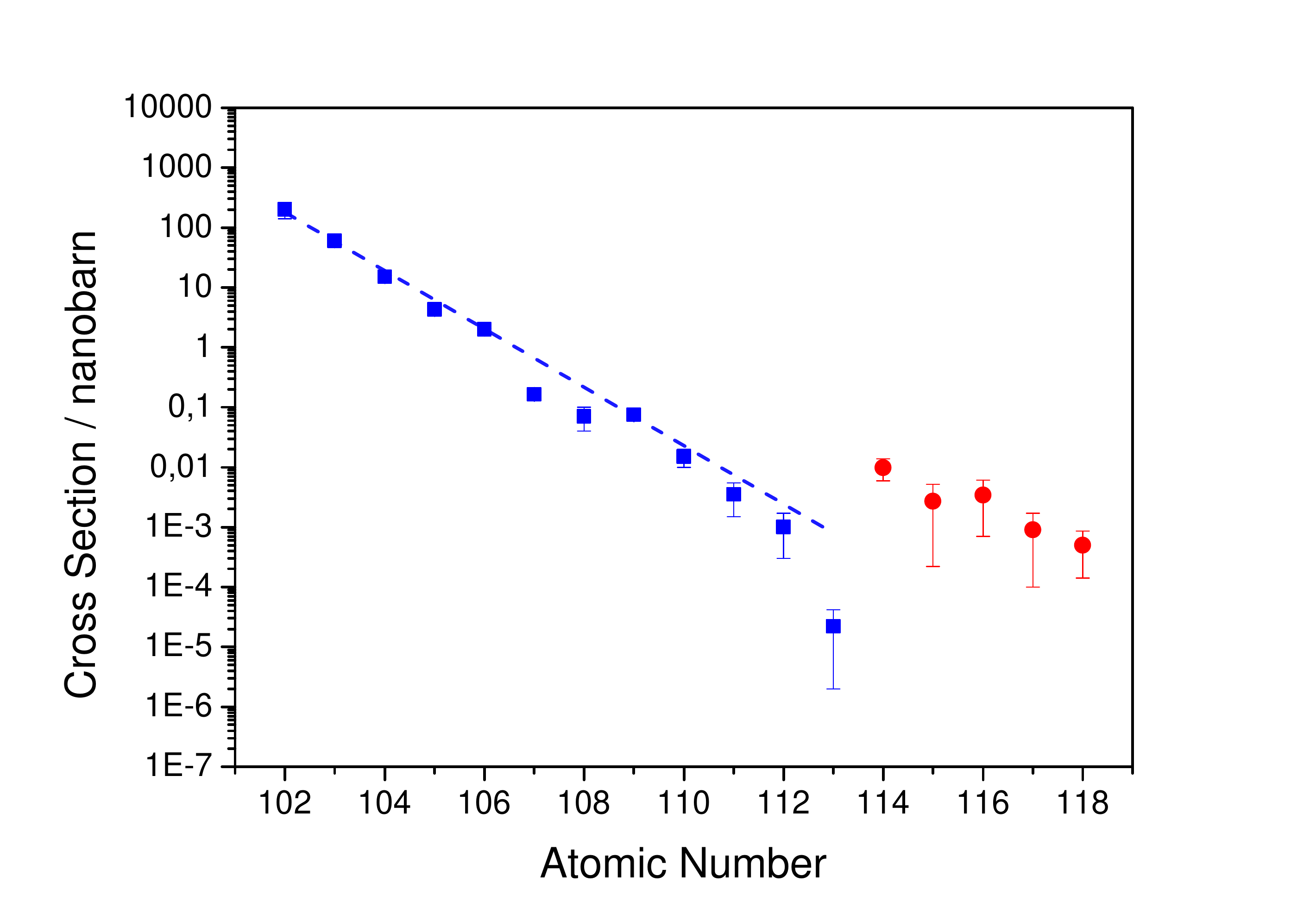}
 \caption{\label{fig:she_cross-section}Maximum cross section for the synthesis of superheavy elements in the one-neutron-evaporation channel in different fusion-evaporation reactions as function of the atomic number $Z$. Data for cold-fusion reactions are shown by the blue squares together with a best fit (dashed blue line). Data for hot-fusion reactions are given by the red dots. The data has been taken from references \cite{Hofmann2000,Oganessian2015,Morita2015a,Muenzenberg2015} and references therein.}
\end{figure}
Figure~\ref{fig:she_cross-section} shows how the cross sections drop exponentially with increasing $Z$ number from a level of close to one microbarn for the synthesis of nobelium\footnote{The cross section for the synthesis of $^{254}$No in the two-neutron evaporation channel is about 2 microbarn.}, Z=102, to the infamously small number of a few femtobarns for the synthesis of element 113 \cite{Morita2004d,Morita2007}. The hot-fusion reactions exploit beams from $^{10}$B -- $^{48}$Ca impinging actinide targets from uranium to californium, and in few cases even einsteinium, to produce more neutron-rich actinides and mainly transactinides. The excitation energy of the compound nucleus is higher, typically around 50~MeV resulting in the evaporation of four to five neutrons. A region of increased cross sections on the level of ten picobarn around $Z=114$ opened the door for the discovery and the investigation of the elements with $Z= 114-118$ \cite{Oganessian2015}.

The fusion products recoil out of the thin target and travel due to momentum conservation forward in the direction of the primary beam. For most of the experimental studies electromagnetic recoil separators are utilized to separate the reaction products from the primary beam in flight and to suppress unwanted reaction products, for example from transfer reactions. Two different types of separators are employed, either vacuum separators such as the velocity filter SHIP at the GSI \cite{Muenzenberg1979,Hofmann2000}, the Recoil Ion Transport Unit (RITU) separator at JYFL \cite{Saren2011}, the Separator for Heavy ELement Spectroscopy (SHELS) at the Flerov Laboratory for Nuclear Research (FLNR) in Dubna, Russia \cite{Rezynkina2015}, the upcoming S$^3$ separator at GANIL \cite{Dechery2016} or gas-filled separators such as the Transactinide Separator and Chemistry Apparatus (TASCA) at the GSI \cite{Semchenkov2008,Gates2011}, the Dubna Gas-Filled Recoil Separator (DGFRS) at the FLNR \cite{Tsyganov1999}, the Berkeley Gas-filled Separator (BGS) at LBNL \cite{Ninov1998,Gregorich2000} and the Gas-filled Recoil Ion Separator (GARIS) at the Institute of Physical and Chemical Research (RIKEN) \cite{Morita1992,Morita2015a}, Japan.

Vacuum separators are based on the principle of the Wien filter with crossed (perpendicular) electric and magnetic fields to separate the particles according to their velocity. Since the kinematics in cold-fusion reactions is such that the velocity of the recoils differs significantly from the projectiles a good primary beam suppression of about $10^{12}$ is achieved. The separation is independent of the charge state.

In gas-filled separators magnetic fields in different configurations of quadrupoles and dipoles are used. The magnets are filled with light gases, mostly He and H, at a low pressure of about 0.5~mbar. In charge changing collisions between the buffer gas atoms and the ion of interest an average charge state with a narrow charge state distribution emerges. The separation is then based on the magnetic rigidity. As a consequence of this charge state equilibrium the transmission efficiency of gas-filled separators may exceed the efficiency of vacuum separators, in particular for asymmetric reactions with actinide targets. The total efficiency that also depends on the acceptance and the kinematics is about 40-80\% for more asymmetric reactions (hot fusion). Gas-filled separators can be quite compact, which may lead to a higher background of unreacted projectiles and scattered beam as well as of transfer products. Furthermore, background from neutrons and gammas at the target position and the beam stop may be higher than for vacuum separators. In contrast to ISOL systems the recoil separators do not feature a significant mass resolving power.

The typical kinetic energy of the recoils is in the range of 10--40~MeV. For laser spectroscopy this requires an additional stage to slow the products down. This is discussed in more detail in section \ref{sec:buffer_gas}. For typical primary beam intensities of the order of $6\times10^{12}$ particles per second typical yields for the nuclides of interest are at most 5--10 atoms per second in the focal plane of the recoil separator for cross sections on the microbarn level.

\subsection{Production in multi-nucleon transfer reactions}

Another nuclear reaction that gives access to more neutron-rich isotopes are deep-inelastic or multi-nucleon transfer reactions. In this approach ultimately a heavy projectile is hitting a heavy target and transfers multiple nucleons instead of completely fusing. Several nucleons are transferred from the projectile to the target. Such reactions have already been studied in the 1980s with respect to the production of neutron-rich superheavy nuclei \cite{Gaeggeler1989,Kratz_2015}. However, applying radio-chemical methods for isotope identification limited the accessible half-life range of the investigated nuclei to a couple of seconds in the best case \cite{Kratz_2015,Schaedel1988}.

Nowadays, physical methods have been applied instead, for example by using electromagnetic spectrometers or the identification by correlated decay chains. For instance, the production of nuclei with $Z\geq92$ has been studied at SHIP \cite{Devaraja2015} and TASCA \cite{Nitto2018} at the GSI in the recent years. In general, the findings of the recent works are largely comparable to the earlier work using chemical methods \cite{Gaeggeler1989}. This mechanism may give access to some previously unstudied, more neutron rich isotopes in the region from uranium up to the elements around $Z=106$ in the near future.

Recent work on laser spectroscopy of multi-nucleon transfer reaction products in the Os isotopes near $N=126$ has been started by the KEK Isotope Separation System (KISS) collaboration \cite{Hirayama2017,Hirayama2020}. The multi-nuclear transfer reaction between a $^{136}$Xe beam and a $^{198}$Pt target has been adopted to produce the nuclides of interest. The reaction products are stopped in an argon-filled buffer-gas cell, neutralized and then laser ionized. 
Another project utilizing multi-nuclear transfer reactions stopped in a gas cell to produce nuclides along the $N=126$ shell is ongoing at Argonne National Laboratory, USA \cite{Savard2020} and may in the future also be exploited for laser spectroscopy.

\section{Selected recent results}
\label{sec:results}

In this section selected examples on recent progress in actinide studies by laser spectroscopy are presented. They reflect the wide range of physics cases as well as the breadth of the methods employed. We sorted the different cases according to rising atomic number.

\subsection{Laser spectroscopy on Ac isotopes at LISOL and TRIUMF}

An extended chain of isotopes was measured for actinium including isotopes at and below the $N$=126 shell closure \cite{Ac_ferrer2017towards,Ac_Granados} and around the region of expected nuclear octupole deformation around $N$=136 \cite{Ac_Verstraelen2019}.

The first study made use of the fusion-evaporation reaction of a $^{22}$Ne beam with a $^{197}$Au target to produce neutron-deficient Ac isotopes at the LISOL facility in Louvain-la-Neuve, Belgium. The reaction products were stopped in a buffer-gas cell where resonance ionization spectroscopy was performed. Heavier Ac isotopes were produced by spallation of a uranium target at the ISOL facility of TRIUMF, where RIS was performed in the hot surface ion source acting as an atom cavity.
    
\begin{figure}[t]
    \centering
    \includegraphics[width=0.8\textwidth]{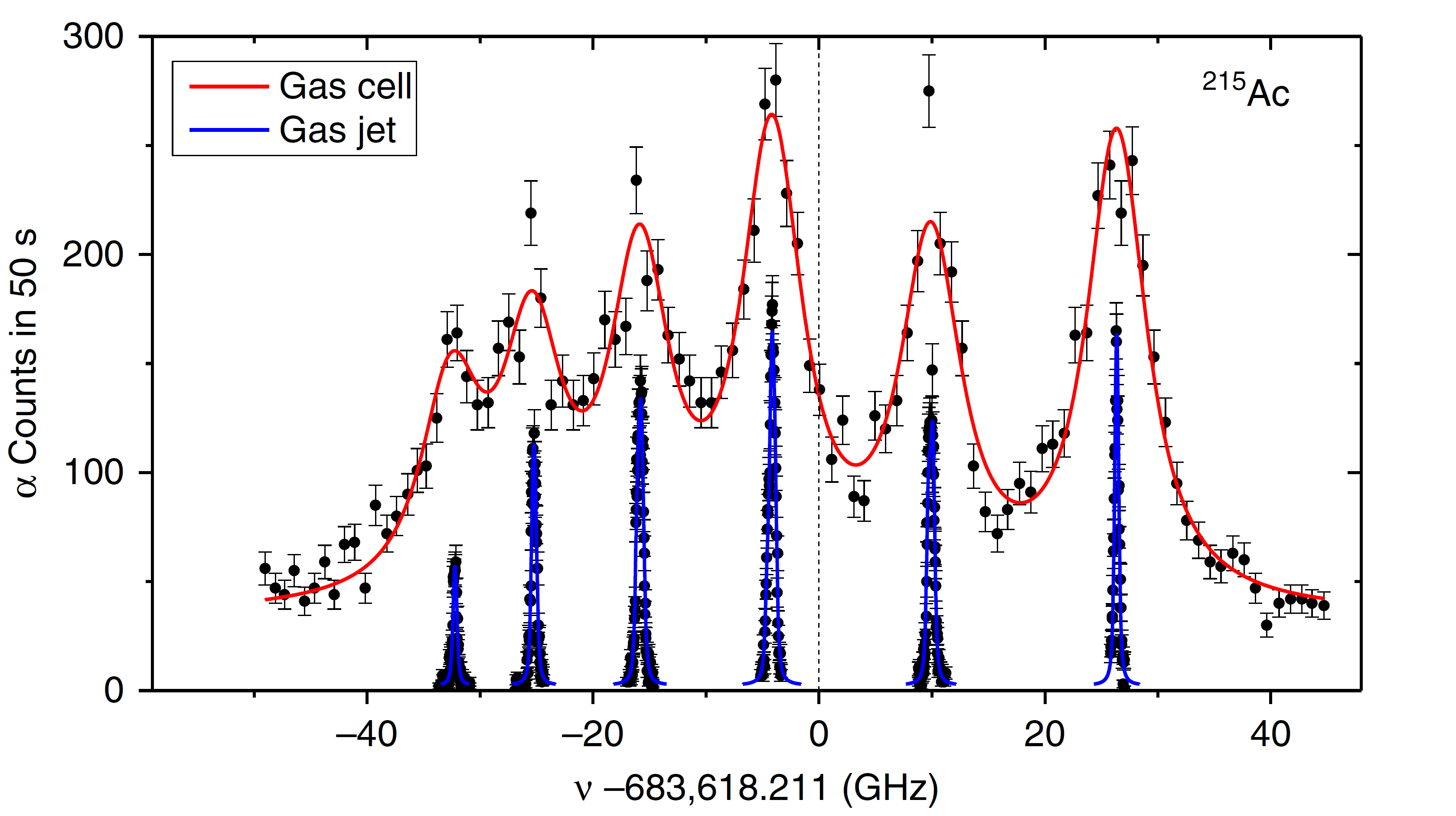}
    \caption{\label{fig:Acspectra}Comparison of the spectrum of $^{215}$Ac measured in-gas cell (red) and in a gas jet (blue), taken from \cite{Ac_ferrer2017towards}.}
\end{figure}
    
The resolution was limited to a few GHz due to collisional broadening for in-gas cell experiments, Doppler broadening for the hot cavity and the laser bandwidths in both cases. A higher resolution of 200-400\,MHz was achieved in RIS-based experiments utilizing injection-seeded lasers \cite{laser_sonnenschein} for $^{225,227}$Ac in an atomic beam \cite{Rossnagel2012} and for $^{214,215}$Ac in a gas jet emerging from a buffer-gas stopping cell \cite{Ac_ferrer2017towards}. 
 
For the laser spectroscopy inside the gas cell a 100\,Hz Excimer-pumped laser system was used \cite{Ac_Granados}, while for the work in the gas jet and in the hot cavity a Nd:YAG-pumped Ti:sapphire laser system operated at a repetition rate of 10\,kHz was employed. 
The ionization scheme in these measurements made use of a first excitation step from the atomic ground state to an excited state at 22\,801\,cm$^{-1}$. The splitting of the excited state dominates the HFS ($A=2105$\,MHz and  $B=110$\,MHz for the reference isotope $^{227}$Ac) which can be resolved in the gas cell as well as in hot cavity environments.
The splitting of the ground state with HFS constants of $A=51$\,MHz and $B=590$\,MHz for $^{227}$Ac could be measured only when applying the in-gas-jet spectroscopy technique.  
The subsequent second transition was driven by laser light of a wavelength of 425\,nm into an auto-ionizing state at 46\,347\,cm$^{-1}$ which could be well saturated with the available laser pulse energies of about $200\, \upmu$J/pulse for the high-repetition rate systems. 

Figure~\ref{fig:Acspectra} shows a comparison of the $^{215}$Ac spectra measured with in-gas cell and in-gas jet techniques.
The in-gas jet spectroscopy results nicely illustrate the gain in resolution compared to those obtained from spectroscopy in the cell. In particular, for hyperfine spectroscopy a high resolution that allows resolving individual hyperfine components is desirable for an unambiguous determination of the nuclear properties from the data.  

The recent Ac measurements, combined with state-of-the-art atomic calculations \cite{Beerwerth2019}, enabled the extraction of the changes in mean-square charge radii and a verification and validation of reported nuclear moments of $^{227}$Ac. The nuclear moments of the other investigated Ac isotopes were inferred based on this reference isotope. 

The obtained charge radii were compared to nuclear calculations, indicating that octupole deformation plays an important role in the observed deformation and raising some questions on the interpretation of the inverted odd-even staggering observed in this region of the nuclear chart \cite{Ac_Verstraelen2019}.

The available yields for the Ac isotopes were still well above 1000 particles per second in all the reported cases. Thus, with an improved data acquisition, we expect more isotopes of Ac to become available in the future despite the refractory character of this element.    
    
\subsection{Pu laser spectroscopy at Mainz and JYFL}
\begin{figure}
    \centering
    \includegraphics[width=0.8\textwidth]{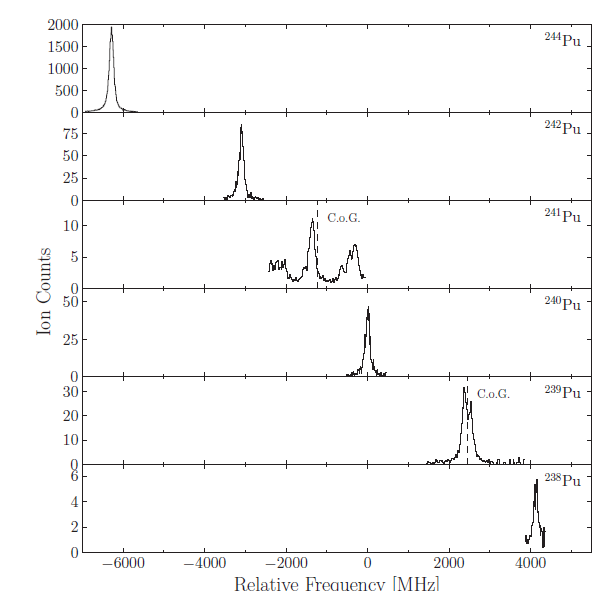}      
    \includegraphics[width=0.78\textwidth]{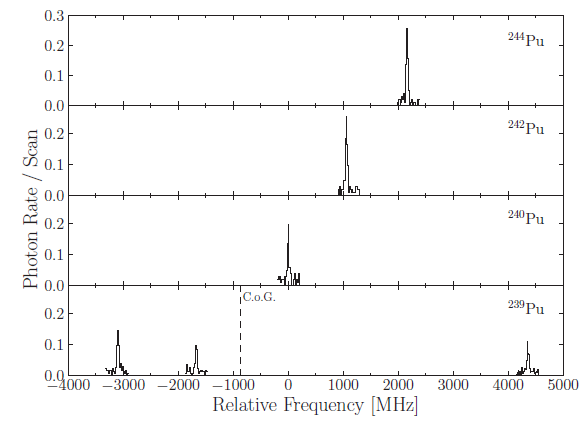}   
    \caption{\label{fig:Puresults} Results from Pu laser spectroscopy by RIS (top) and CLS (bottom) \cite{Pu_voss2017}.}
\end{figure}

A laser spectroscopic study with a high spectral resolution was performed on the long-lived plutonium isotopes $^{238-242,244}$Pu \cite{Pu_voss2017}.
Here, the technique of collinear laser spectroscopy (CLS) on a fast ion beam with fluorescence detection as well as RIS in the atomic beam emerging from a hot cavity were employed.
 At JYFL, a fast beam of plutonium for the CLS beam line was produced from a heated filament inside a buffer-gas cell of the IGISOL front-end using resonant laser ionization and final acceleration of the created laser ions in an electrostatic field gradient.

For these studies the 363\,nm-transition from the 5f$^6$7s\,$^8$F$_{1/2}$ ground state to a $J$=1/2 excited level at 27\,523\,cm$^{-1}$ in singly-ionized Pu was investigated.
Locking a cw-laser to a fixed frequency and tuning the velocity of the fast-ion beam allowed performing spectroscopy on the isotopes $^{239,240,242,244}$Pu at a spectral linewidth of about $40\,$MHz as shown in the bottom panel of figure\,\ref{fig:Puresults}.

RIS measurements in the emerging atomic beam incorporating long-lived Pu isotopes from a hot cavity were performed at Mainz using a dedicated off-line setup comprising a quadrupole mass spectrometer. Here, two optical transitions in neutral Pu atoms were investigated, one at 385\,nm starting from the 5f$^6$7s$^2$\,$^7$F$_0$ ground state to the 5f$^5$6d$^2$7s\,$J$=1 excited level and a second at 388\,nm from the 5f$^6$7s$^2$\,$^7$F$_1$ thermally excited level at 2203.6\,cm$^{-1}$ to the 5f$^6$7s7p\,$J$=2 excited level.
Ionization was obtained by a subsequent, second transition into an auto-ionizing state using laser light at 407\,nm in the first case and 436\,nm in the latter.

These measurements allowed studying two optical transitions in $^{238-242,244}$Pu at a spectral linewidth of about $150\,$MHz.  
The laser system employed two pulsed Ti:sapphire lasers with a 10\,kHz repetition rate, including an injection-locked laser with a bandwidth of 13-20\,MHz that was used for scanning the spectroscopic transitions \cite{laser_sonnenschein}.
An example of the results obtained with RIS is shown for the case of the 388\,nm-transition in the top panel of figure\,\ref{fig:Puresults}.

The study of Pu isotopes with different techniques at a similar precision enabled a thorough comparison of these complementary methods, which agree with each other at a level that is about one order of magnitude better than the stated literature values \cite{Gerstenkorn1987, Angeli2013}.
Nevertheless, the precision of the final changes in the mean square charge radii is presently still limited by the calibration to the isotope shift, which so far is obtained from independent measurements of absolute charge radii \cite{Zumbro1986}.

\subsection{Spectroscopy of $^{229}$Th}
\label{sec:thorium}

The investigation of $^{229}$Th and its isomer, $^{229m}$Th, is continuing to attract attention for more than four decades, since the first days when Kroger and Reich indicated the existence of a nuclear ground-state doublet of extremely low energy splitting~\cite{Kroger1976}. In the recent years, the topic developed a genuine rally to characterize the isomer and pin down its energy with highest precision~\cite{Bilous:2017,Jeet:2015,Kazakov:2014,Wense2016,Seiferle:2017}. Several research papers appeared suggesting this splitting to be below $10\,$eV~\cite{Beck:2007,Seiferle:2019,Masuda:2019}, which bears the potential of making a nuclear transition accessible for laser probing for the first time.

The thorium isomer $^{229m}$Th, thus, provides the best candidate for the development of a new generation of optical clocks. Based on an extremely narrow nuclear transition, which is believed to be insensitive to external fields, such a clock could outperform current atomic clocks in precision by up to about one order of magnitude. In addition, new routes in applied and fundamental research would open up by better knowledge of the energy splitting~\cite{Flambaum:2006,Thirolf2019a,Thirolf2019b}.

Only recently, Seiferle {\it{et al.}}~\cite{Seiferle:2019} reported a value of $8.28\pm0.17\,$eV for this splitting, narrowing down the energy uncertainty to ranges amenable for laser scans at wavelengths around $150$~nm. Working towards this goal, diverse experimental approaches are currently considered, ranging from direct nuclear excitations~\cite{Verlinde:2019,Wense2017} to exploiting electronic bridge processes~\cite{Porsev2010}.

For isomer diagnostics utilizing the nuclear-electronic double resonance method, Th$^{3+}$ with a relatively simple electronic structure has been considered~\cite{Peik:2003}. The ion exhibits closed two-level and three-level systems for laser probing and cooling at the same time, see figure~\ref{fig_scheme}(a).

\begin{figure}
 \centering
 \includegraphics[width=0.8\textwidth]{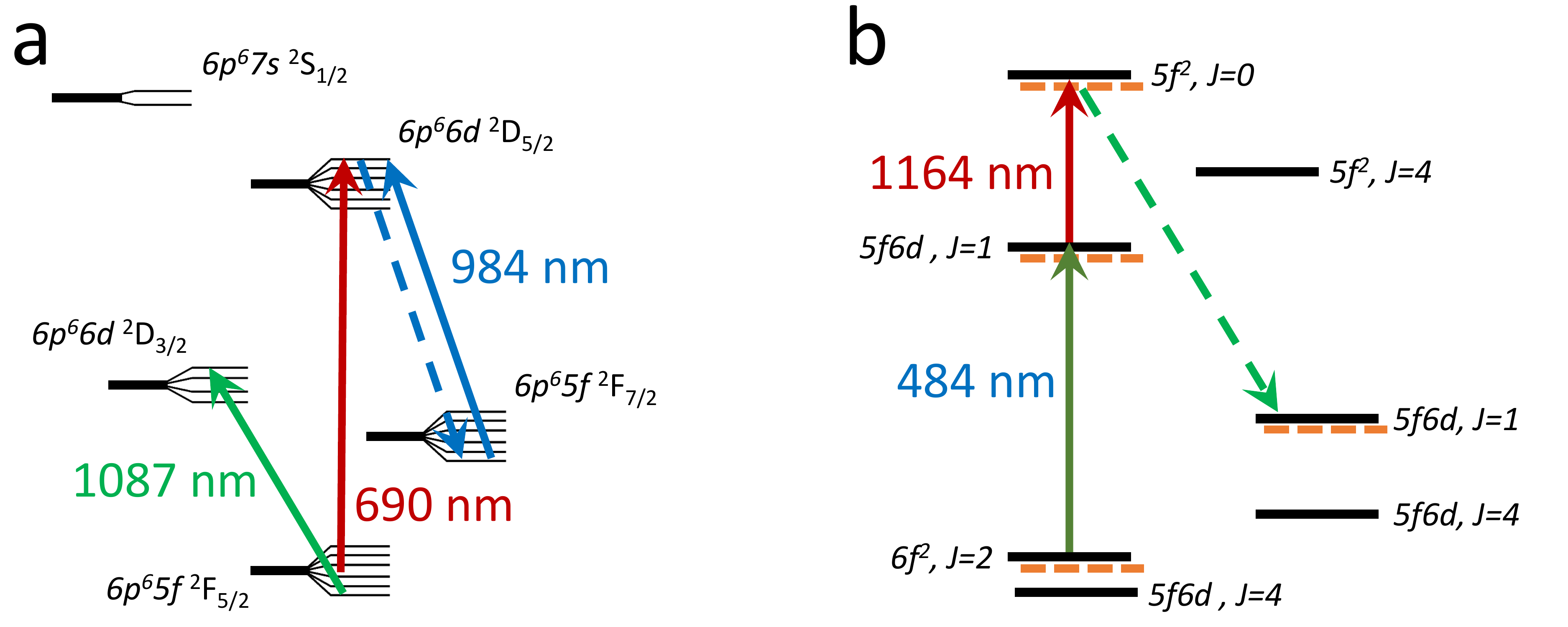}
 \caption{\label{fig_scheme} Excitation schemes used in HFS of $^{229}$Th$^{3+}$ in the ground state~\cite{Campbell:2011} (a) and of $^{229}$Th$^{2+}$ in ground- and isomeric state~\cite{Thielking:2018} (b). Solid (dashed) arrows indicate laser excitations (fluorescence). Stars indicate electronic states of the nuclear isomer. Schemes adapted from Refs.~Refs.~\cite{Campbell:2011,Thielking:2018}.}
\end{figure}

In addition, most systematic frequency shifts can be eliminated when monitoring isomeric excitations with an occupied $^2$S$_{1/2}$ state of this ion. For more details we refer to other reviews on this field~\cite{Peik:2015,Thirolf:2019}.
Successful hyperfine spectroscopy of $^{229}$Th$^{3+}$ in the nuclear ground state has been reported in 2011 by Campbell {\it{{\it{et al.}}}~}\cite{Campbell:2011}. The ion ensembles were created by laser ablation from a $^{229}$Th(NO$_3$)$_4$ target. The excitation scheme used is depicted in figure~\ref{fig_scheme}(a). Multiple optical fields were used to repump the $^2$D states from the $^2$F$_{5/2}$ and $^2$F$_{7/2}$ manifolds. By creating Wigner crystals at mK temperatures inside a linear Paul trap, Doppler broadening was reduced to a few-MHz level. For the three shown transitions, isotope shifts relative to the $^{232}$Th reference were determined. In addition, a spectroscopic quadrupole moment of $Q_s=3.11(16)\,$eb was obtained.

Only recently, in 2018, hyperfine spectroscopy of the isomer was performed using $\alpha$-recoil ions from a $^{233}$U sample of about $290\,$kBq activity~\cite{Thielking:2018}. The decay branch into $^{229m}$Th is about $2$\%. Prior to spectroscopy, the recoil ions were thermalized in a helium-filled stopping cell~\cite{Wense2015,Thirolf:2019} and extracted and confined in a radiofrequency trap at room temperature. The measurements were performed on Th$^{2+}$ because of its convenient electronic structure, which enables hyperfine spectroscopy with diode lasers with background-free fluorescence detection in the visible spectral range~\cite{Thielking:2018}. The corresponding excitation scheme is shown in figure~\ref{fig_scheme}(b). Starting from the $6f^2$ level, two-step laser excitations with co- and counter-propagating laser beams of $484\,$nm and $1164\,$nm wavelengths enabled dentification of all nine resonances of the nuclear ground state and to observe seven out of eight expected resonances of $^{229m}$Th$^{2+}$.

A detailed analysis of the spectra revealed isomeric shifts, which correspond to a slightly increased mean-square charge radius of the isomer by $0.012(2)\,$fm$^2$. The magnetic dipole moment was found to be $-0.37(6)$ nuclear magneton units, while the spectroscopic quadrupole moment amounts to $Q_s =1.74(6)\,$eb. The corresponding intrinsic quadrupole moment was found to be nearly the same as measured for the nuclear ground state, which implies similar prolate-shaped nuclear charge distributions for both states.

\subsection{Quantum chaos in the atomic spectrum of protactinium}
\label{sec:protactinium}

An example nicely illustrating the complexity of atomic structure investigations in the region of actinides, where several orbitals are accessible, is the work from Naubereit {\it{et al.}} \cite{Naubereit2018a}.
In their studies the atomic structure of the element protactinium was probed by laser spectroscopy using a laser system with a repetition rate of 10\,kHz in a hot cavity with an off-line sample at the University of Mainz.
In total, about 1500 previously unknown high-lying atomic levels were identified \cite{Naubereit2018}, mainly in the energy region from 48\,000\,cm$^{-1}$ to 50\,000\,cm$^{-1}$ around the first ionization potential of 49\,100(100)\,cm$^{-1}$ \cite{Wendt2014}.
The observed high level density results from the five valence electrons in the Pa atom which can access the 5f, 6d, 7s and 7p orbitals allowing for a multitude of electron configurations.

Based on theoretical predictions indicating a potentially chaotic behavior of the atomic structure \cite{Viatkina2017} these experimental levels were then investigated.
Using statistical methods and simulation studies indicated that from the neighbor spacing and the level density for this atom the level distribution does no longer match classically integrable systems but rather shows an intrinsically fully chaotic behavior already at energies well below the first ionization potential.

\subsection{Laser spectroscopy of nobelium isotopes at SHIP at the GSI}
\label{sec:nobelium}

The RADRIS experiments at the GSI extended laser spectroscopy to the then unexplored element nobelium, $Z=102$, the common gateway to the region of the heaviest elements. Different nobelium isotopes are conveniently produced in cold-fusion reactions of $^{48}$Ca with different lead targets. The cross section for the production of the isotope $^{254}$No is the largest one in the entire region above fermium with about two microbarn. This results in typical production rates on the order of 15 particles per second under standard experimental conditions. 

From the atomic physics point of view the nobelium atom is favorable for laser spectroscopy due to its relatively simple ground state configuration [Rn] $7s^25f^{14}$. The combination of both aspects make $^{254}$No the natural choice for the first foray into this region of the nuclear chart. Theoretical predictions for the strongest atomic ground-state transitions in the nobelium atom from different groups using different models were available \cite{Fritzsche2007,Indelicato2007,Borschevsky2007a}. These calculations were motivated by the experimental plans of Backe {\it{et al.}} \cite{Backe2007} who searched the spectral range given by the then available predictions and their uncertainties in a first experiment performed at GSI starting in 2007 \cite{Laatiaoui2014b} without finding any transition. Later on additional values from other groups were published \cite{Liu2007,Dzuba2014} and shifted the search region. 

A partial level scheme of the nobelium atom showing the strongest transition from the ground state is given in figure~\ref{fig_no-trans}. The ionization potential of nobelium had been predicted to be about 6.6\,eV.
Based on the new predictions and after implementing a couple of technical and methodological improvements \cite{Lautenschlaeger2016} the level search was resumed and eventually a first transition was identified. 

\begin{figure}
    \centering
    \includegraphics[width=0.8\textwidth]{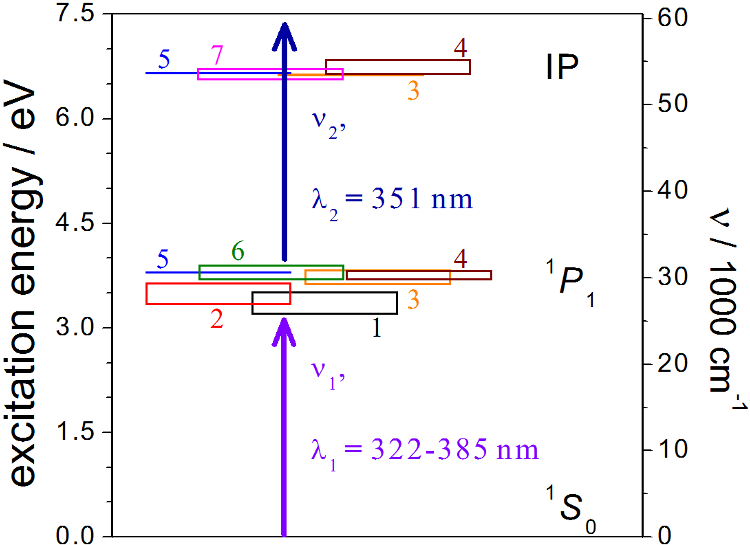}
    \caption{\label{fig_no-trans} Theoretical prediction for some levels in atomic nobelium. The values are taken from 1,2: \cite{Fritzsche2005},3:\cite{Borschevsky2007a},4:\cite{Dzuba2014},5:\cite{Liu2007},6:\cite{Indelicato2007} and 7:\cite{Sugar1974}}.
\end{figure}

Laatiaoui {\it{et al.}} reported the identification of the strong $^1$S$_0-^1$P$_1$ transition in nobelium at 29961.5\,cm$^{-1}$ using the RADRIS technique \cite{Laatiaoui2016}. They applied a two-step excitation scheme with a resonant first excitation step and a non-resonant second step into the continuum. The first step was provided by a pulsed dye laser pumped by an excimer laser. The pulse width of about 18~ns and the repetition rate of 100~Hz were suitable for in-gas cell laser spectroscopy where the typical drift times in the gas are in the millisecond regime. The second step was provided by an excimer laser (at 351~nm) with an energy of about 20-50\,mJ per pulse. The saturation characteristics of the $^1$S$_0-^1$P$_1$ transition was determined experimentally to verify that based on the Einstein coefficient for absorption the lifetime of the state is in agreement with the calculations for the $^1$P$_1$ state of only about 2~ns. The overall efficiency using a tantalum filament was about 6--12\,\%.

After finding the first atomic level in good agreement with theoretical predictions the measurements were extended to search for Rydberg states in a two-step excitation scheme using two pulsed dye lasers. The first excitation step was fixed on the found resonance and the second step was scanned. Due to the short lifetime of the $^1$P$_1$ level the temporal synchronization was crucial. As reported by Chhetri {\it{et al.}} several Rydberg states were identified \cite{Chhetri2018}. However, it was noticed that the identified Rydberg series originate from different intermediate levels. From the convergence and the corresponding series limits the first ionization potential of nobelium was determined to be $6.626 21(5)\,$eV \cite{Chhetri2018}. Their result was in excellent agreement with the theoretical predictions and with a value of moderate accuracy that was recently reported by Sato {\it{et al.}} \cite{Sato2018} based on the temperature dependence of surface ionization.

\begin{figure}
    \centering
    \includegraphics[width=0.7\textwidth]{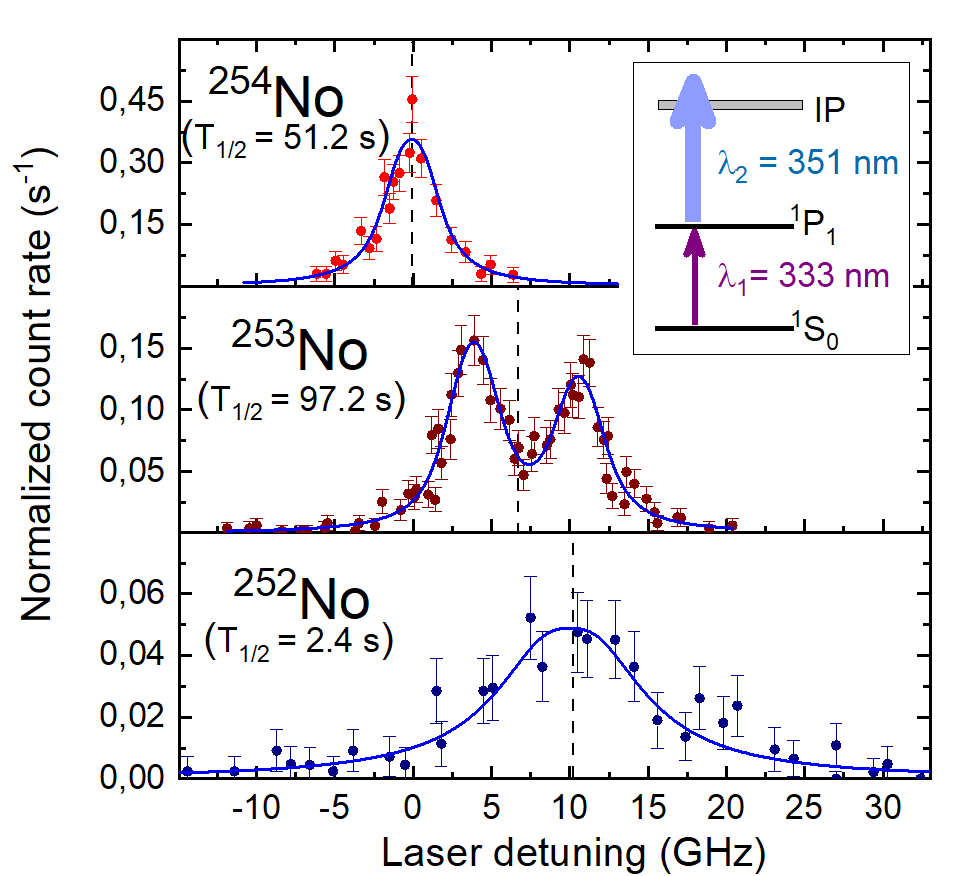}
    \caption{\label{fig_nobelium} $^1$S$_0-^1$P$_1$ transition in the nobelium isotopes $^{252-254}$No. The solid lines show a fit of a rate model equation to the experimental data. The transition in $^{253,254}$No was recorded with an etalon in the laser to reduce the linewidth. Figure adapted from Ref.~\cite{Raeder2018}.}
\end{figure}

A peculiarity of the nobelium atom was observed in this context. The buffer-gas environment led to collision-induced quenching populating the metastable $^{3}$D levels. According to theory, these levels lie only a few hundred wavenumbers below the $^1$P$_1$ level and can thus be populated in this process. The non-resonant second excitation step at 351~nm has sufficient energy to ionize atoms even from the metastable states. From the measured response to a delayed ionization step it was possible to disentangle the different excitation pathways and to assign the different Rydberg series. This was possible based on the short lifetime of the $^1$P$_1$ state of a few nanoseconds in contrast to the lifetime of hundreds of nanoseconds for the close-lying metastable $^{3}$D states. As a byproduct it was possible to obtain the energy difference between the $^3$D$_3$ and $^1$P$_1$  level considering the input from the theoretical calculations by Borschevsky {\it{et al.}} \cite{Borschevsky2007a}. 

Once the atomic properties in the nobelium atom were settled the measurements focused on the nuclear properties and the extension to additional nobelium isotopes. The isotopes $^{252-255}$No were produced in the one and two-neutron evaporation channels of the fusion evaporation reactions of a $^{48}$Ca beam with $^{206-208}$Pb targets. However, the production cross sections drop compared to $^{254}$No by up to a factor of forty-five as shown in table~\ref{tab:cross-section}.

\begin{table}[htp]
\caption{Production of the isotopes of interest and some key properties. $^{255}$No can be either produced directly or indirectly via the electron capture decay branch of $^{255}$Lr. The latter has a lower cross section but avoids background from the production of $^{254}$No. For further details see text.}
\label{tab:cross-section}
\begin{center}
\begin{tabular}{|c|c|c|c|}
\hline
Nuclide & Half-life (ground state) / s & Reaction & Cross section / nb \\
\hline
$^{252}$No & 2.44 & $^{48}$Ca($^{206}$Pb, 2n) & 400 \\
$^{253}$No & 97 & $^{48}$Ca($^{207}$Pb, 2n) & 1000  \\
$^{254}$No & 51 & $^{48}$Ca($^{208}$Pb, 2n) & 1800 \\
$^{255}$No & 211 & $^{48}$Ca($^{208}$Pb, 1n) & 140 \\
$^{255}$No & 211 & $^{48}$Ca($^{209}$Bi, 2n), EC & $\approx$40 \\
\hline
\end{tabular}
\end{center}
\label{default}
\end{table}

Despite the decreasing yield it was feasible to identify the $^1$S$_0 - ^1$P$_1$ transition also in $^{252-254}$No to determine the isotope shift \cite{Raeder2018}. The obtained spectra are shown in figure~\ref{fig_nobelium}. The linewidth was limited by the Doppler and pressure broadening and the laser bandwidth to about 3--5 GHz. The shift of the line center is clearly visible. The spectral resolution in the case of $^{252}$No was lower due to an increased laser bandwidth, but in principle it can be measured with the same resolution as the other isotopes. In $^{253}$No with a nuclear spin parity of $9/2^+$ previously established from nuclear spectroscopy experiments \cite{Asai2015} the hyperfine splitting is observed. With the achievable spectral resolution two out of three expected hyperfine components were resolved under these conditions. Nonetheless, the hyperfine parameters $A$ and $B$ were obtained from a fit of the expected line shape to the data points. From these parameters in combination with different atomic calculations the magnetic dipole moment and the spectroscopic quadrupole moment were determined \cite{Raeder2018,Porsev_2018}.

\subsection{Nuclear charge radii in the actinide region from laser spectroscopy}
\label{sec:act-radii}

Nuclear charge radii derived from laser spectroscopy attract a strong interest as they give access to the nuclear size and shape. One aspect is to track the evolution of the nuclear size and shape to identify regions where the deformation changes. In the region of the actinides and transactinides the nuclei at the major shell closures at $N=126$ and the predicted one at $N=184$ are spherical. In between, regions of well-deformed nuclei with oblate and prolate deformation occur. The majority of the known superheavy nuclei are prolate. For example, the deformation for the nobelium nuclei around $N=152$ is $\beta \approx 0.2$ \cite{Asai2015,Raeder2018}. 

Another motivation is the search for rather exotic phenomena that are predicted to occur in certain regions far-off stability. Famous examples are halo nuclei \cite{Tanihata_1985} and octupole deformed nuclei \cite{Gaffney2013}. A specific feature that has fascinated physicists is the occurrence of so-called bubble nuclei, or more precisely nuclei with a central depression in the nucleonic density. This phenomenon has long been predicted, see for example these works \cite{Davies1972,Campi1973,Decharge2003,Schuetrumpf2017} and references therein, but is experimentally difficult to observe. There are two different types of case for which a depression may occur. In light to medium-mass nuclei as for example $^{34}$Si these exotic shapes are ascribed to the sensitivity to nuclear shell effects \cite{Mutschler2016,Duguet2017}. In very heavy nuclei the central depression is due to the electrostatic repulsion that drives protons out of the center of the nucleus towards the surface \cite{Schuetrumpf2017}. 

\begin{figure}
    \centering
    \includegraphics[width=0.8\textwidth]{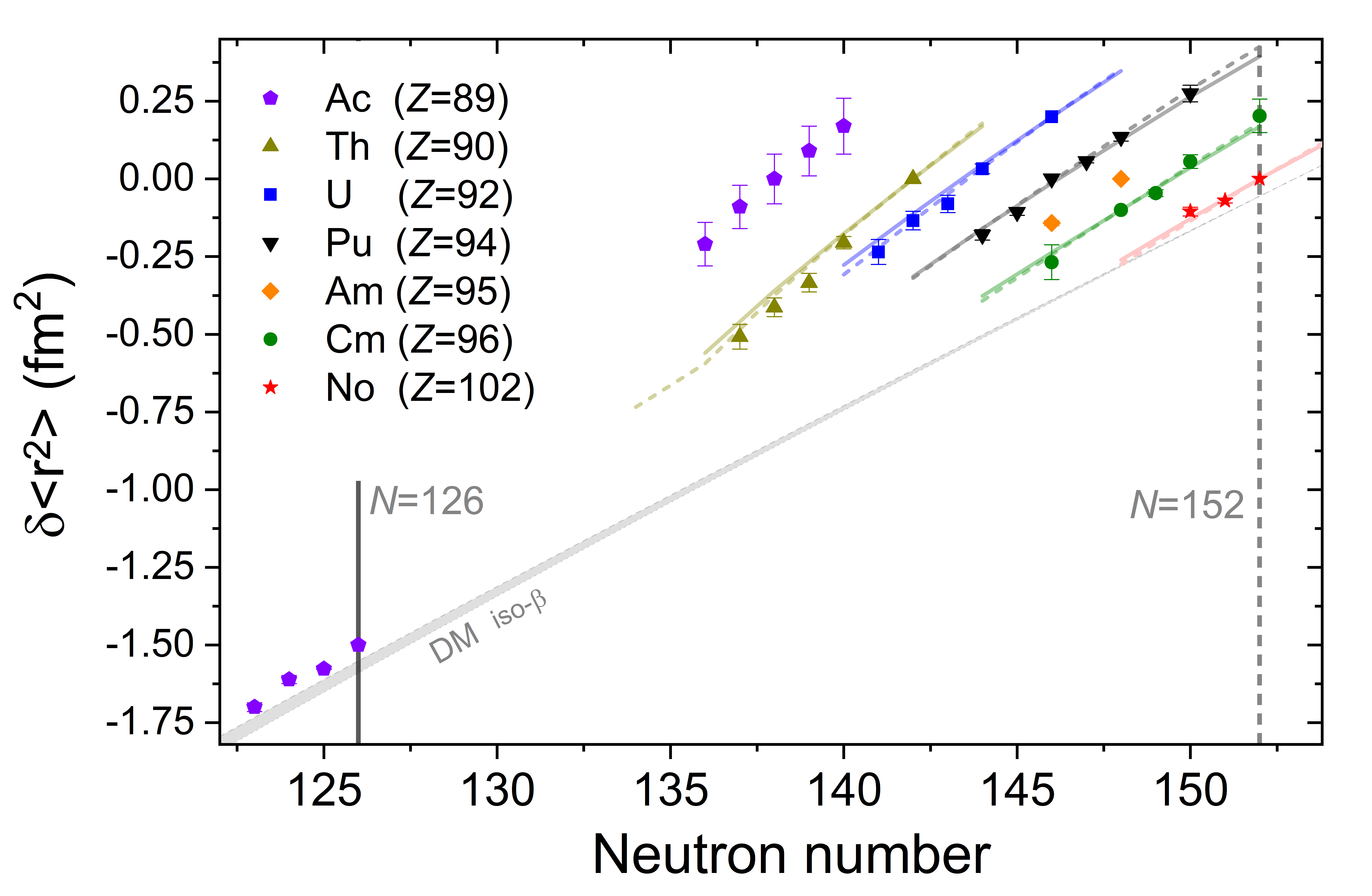}
    \caption{\label{fig_radii} Changes in the mean square charge radii based on laser spectroscopy data for the Ac to No, date taken from \cite{Ac_Verstraelen2019,Ac_Granados,Angeli2013}. Theoretical predictions by an energy density functional approach are given by the lines. For orientation the iso-beta line for a droplet model (DM) is also shown. Vertical lines represent nuclear-shell or sub-shell closures. Figure adapted from Ref.~\cite{Raeder2018}.}
\end{figure}

The changes in the mean square charge radii in the actinides computed from the measured isotope shifts are summarized in figure~\ref{fig_radii} for the isotopic chains of elements from actinium to nobelium in the range $N=123-152$. Thanks to the latest results discussed above a more complete picture starts to emerge. In the plotted region the charge radii increase more or less linearly for all isotopic chains, but with slightly different slopes. This confirms the prolate deformation with $\beta \approx 0.2-0.3$ for most of the measured nuclides. There are presently no data beyond the deformed shell closure at $N=152$.

Along with the experimental data theoretical predictions from a nuclear model based on a Skyrme energy density functional are shown \cite{Schuetrumpf2017,Raeder2018,Reinhard2017}. All the experimental data are remarkably well described by this functional that has neither been adapted specifically for the actinide region nor for the description of charge radii. In fact, it has been rather challenging to describe the nearly identical charge radii of the Ca isotopes $^{40}$Ca and $^{48}$Ca despite the large difference in neutron number, which could not be reproduced in theoretical calculations for a while. Besides an earlier description by a shell-model approach \cite{Caurier2001}, only recently it was possible to describe the data using a Fayans functional \cite{Fayans2000,Ruiz2016,Reinhard2017,Miller2019}. Other theoretical approaches for calculations of charge radii can be found in \cite{Reinhard2017,Ekstroem2015}.

The model that describes the data for the charge radii in the actinides in general and in nobelium particularly well
predicts a central depression in the proton distribution of $^{254}$No of about 10\% \cite{Raeder2018}. The central depression in heavy nuclei has been theoretically studied by several groups. A recent analysis was performed for example by Schuetrumpf {\it{et al.}} \cite{Schuetrumpf2017}. The isotope shift analysis in nobelium supports the model calculations and thus provides indirect evidence for this effect.

One of the consequences is that in the future a more sophisticated parametrization of the charge distribution inside the nucleus will be required in atomic calculations of the isotope shifts. One approach, which accounts for relativistic corrections in the field isotope shifts, has been recently suggested by Flambaum and Dzuba \cite{Flambaum2019}.
 
It will certainly be interesting to investigate the central depression in heavier nuclei, where the effect is predicted to be even stronger. It is also desirable to apply other established techniques such as electron scattering to some exotic radionuclides to look for a signature of such effects. However, experimental limitations such as low production rates and short half-lives make such measurements very challenging. Other measurements on long-lived nuclei such as $^{248}$Cm may be obtained in the near future though and may then provide absolute charge radii to serve as references in the isotope shift analysis from laser spectroscopy. 

\subsection{Ionization potentials of the actinides}
\label{sec:IP}

\begin{figure}
    \centering
    \includegraphics[width=0.8\textwidth]{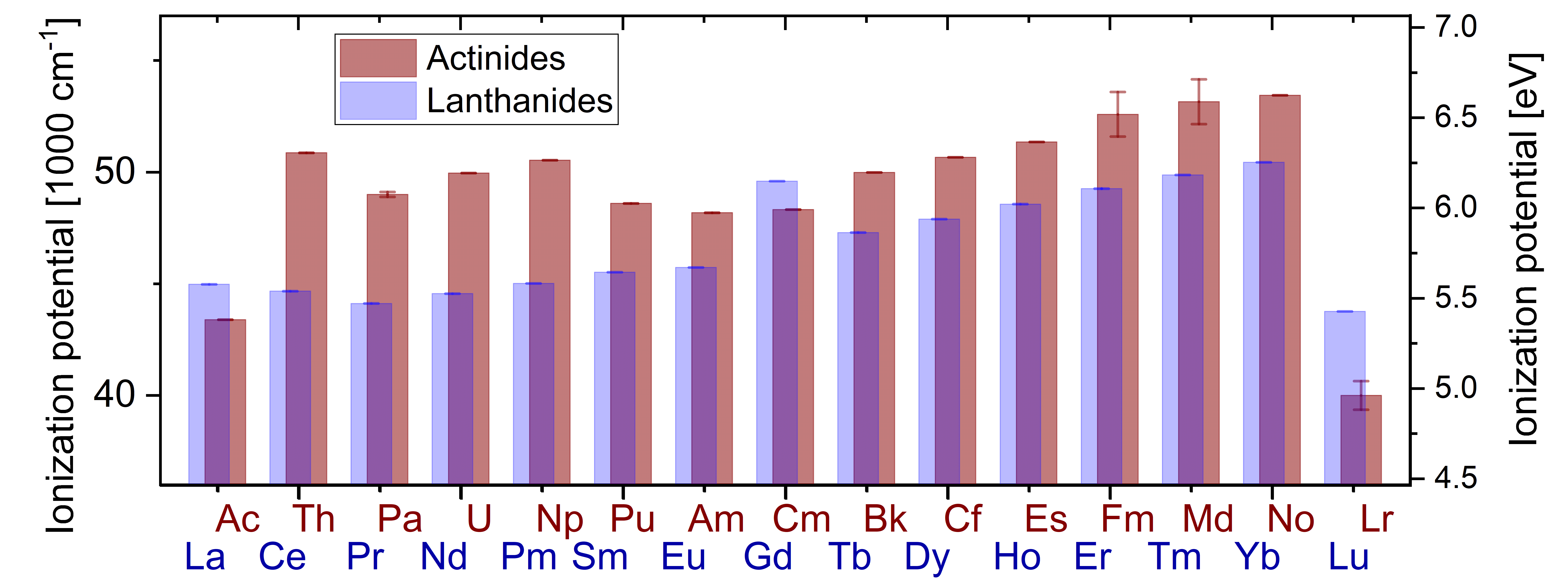}
    \caption{\label{fig_ionizationpotential} First ionization potentials of the actinide series from Ac to Lr, (data taken from \cite{Sato2015,Sato2018,Waldek:2001,Koehler:1997,Rossnagel2012,Wendt2014,Chhetri2018}), the general trend is similar to the lanthanide series (data from\cite{NIST:2018} ). As an effect of the closed atomic shells Lu and Lr show a clear drop in the IP compared to the lighter elements.}
\end{figure}

An important atomic property that determines the chemical behavior of an element is its first ionization potential (IP), the energy needed to remove the weakest bound electron from the atomic shell.
An overview on the ionization potentials for the lanthanide and actinide series was given by Wendt {\it{et al.}} \cite{Wendt2014}, but at that time the IP for the heavy actinides above Es were not yet measured. 
While the actinides up to Es were determined in the 1990s by laser spectroscopy utilizing field ionization with variable electrostatic fields \cite{Waldek:2001,Koehler:1997}, the value for actinium was refined in 2012 by Ro{\ss}nagel {\it{et al.}} \cite{Rossnagel2012} using Rydberg convergence in a hot cavity environment.
Only recently the ionization potentials for the remaining heavy actinides was measured completing the actinide series as shown in figure\,\ref{fig_ionizationpotential}.

A first study by Sato {\it{et al.}} \cite{Sato2015} in 2015 used the ISOL system at JAEA to perform surface ionization of Lr which was produced in a fusion reaction and transported by aerosols in a carrier gas.
From the obtained surface ionization efficiencies in comparison to other lanthanide elements and with the help of atomic calculations to determine the level densities the ionization potential for Lr was derived to be 4,96\,eV or about 40\,000\,cm$^{-1}$.
This work was continued and in 2018 also the ionization potentials for Fm, Md and No were derived with the same technique, limited by the lower ionization efficiencies of these elements with higher IPs \cite{Sato2018}.
The accuracy of the method is about 1000\,cm{$^{-1}$ in the case of Fm and Md and about 400\,cm$^{-1}$ for Lr, which is still significantly less precise compared to the uncertainty from laser spectroscopic investigations.
In nobelium the ionization potential was measured with high precision using the RADRIS technique to be 53\,444.0(4) cm$^{{-1}}$ \cite{Chhetri2018}, see Sec.~\ref{sec:results} above.
Albeit the ionization potentials are now measured experimentally for all actinide elements, the precision is still limited for Pa, Fm, Md and Lr, which remain subjects of future laser spectroscopic investigations.

The main conclusion that can be drawn from the available data is that the actinide series is indeed terminated by lawrencium. Furthermore, one sees that the trend of the IP in the lanthanides and actinides is very similar, but the trend in the IP of the actinides is more irregular. 
This is because relativistic effects are even more pronounced in the actinides, featuring many close-lying levels, which lead to these irregularities.

\section{Perspectives}
\label{sec:perspectives}

Besides the recent laser spectroscopy achievements in the region of the heavy elements discussed above, still many open questions and challenging measurements remain. The general goal is to extend the measurements to ever more nuclides. This requires additional work on increasing the overall efficiency of the methods. At the same time, it is crucial to increase the spectral resolution, in particular for hyperfine spectroscopy. 
 
Nonetheless the future perspectives are largely governed by the production mechanisms and the corresponding yields. In the near future the dominant production scheme will remain to be fusion-evaporation reactions. To progress beyond nobelium and lawrencium the experiments have to deal with nuclides produced with cross sections on the level of tens of nanobarns or less and an increasing refractory character. For present-day accelerators these cross sections correspond to typical rates of a few particles per minute entering the setup (prior to stopping in the gas). Next-generation accelerators with powerful ion sources promise an increase in primary beam intensities by an order of magnitude to about $6 \times 10^{13}$ particles per second \cite{Barth2017a,Dmitriev2016,DROUART2009}. In combination with improved targets that can handle the high beam intensity and optimized recoil separators this may result in ten times higher yields than available today. From the production viewpoint this will enable experiments and more detailed studies of the elements up to about Sg, $Z=106$, in the next couple of years. However, this still constitutes only a moderate push for laser spectroscopy. In order to make a move towards higher $Z$ the total number of atoms required to obtain a certain observable has to be decreased by introducing novel methods. 

\subsection{Off-line measurements of heavy actinides up to Fm}

Isotopes of some heavy actinides are produced in nuclear reactors as discussed in section~\ref{sec:basics}. This also includes minute quantities in the scale of picogram for Fm and nanogram to microgram for Es isotopes. 
To use these samples for laser spectroscopy efficient and highly sensitive techniques are required to account for the low number of available atoms and the low abundances.

Hot cavity resonance ionization spectroscopy reaching high total efficiencies are in principle suitable for measurements utilizing picograms of actinide samples. The requirements have a considerable overlap with trace detection techniques which call for high efficiency and selectivity in conjunction with a low background \cite{Raeder2019b,Liu2020}.
A typical setup consists of a high-efficiency mass separator, ideally employing a sector field magnet, with a hot cavity atom source as well as a high-repetition rate laser system. The requirements for the mass resolving power are moderate as usually only isotope separation is required due to the element selectivity provided by the laser excitation.

With respect to the laser system requirements similar to the discussion in \cite{Raeder2020} dye laser systems as well as Ti:sapphire laser systems offer the possibility for the required high powers as well as the option for narrow bandwidth operation.
For high resolution, RIS in the emerging atom beam with a confinement from a radiofrequency quadrupole structure \cite{Heinke2016} or the laser spectroscopy in a supersonic gas jet \cite{Ac_ferrer2017towards,Raeder2016} offer the required sensitivity and efficiency.

In addition to the spectroscopic setup and the laser system also a profound knowledge of the chemical separation of different actinides as well as of their chemical behavior in general is important for an efficient use of these samples.
For example, at the University of Mainz collaborations of nuclear chemists and atomic physicists have performed experiments on the heavier actinide elements in the past measuring their ionization potential \cite{Koehler:1997,Erdmann:1998}.

Recently, off-line measurements of long-lived actinides have been extended at the RISIKO separator of the University of Mainz. In a comprehensive program several new measurements have been performed on long-lived Pu, Cm, and Cf isotopes and even measurements on $^{253-255}$Es and $^{257}$Fm have recently been achieved. All the measurements used a hot-cavity approach and 10~kHz-Ti:Sa laser systems with frequency doubling. The data are presently still under analysis and will be published in the near future. New and improved data on atomic and nuclear properties are expected from these studies. Additional measurement campaigns are planned some of which are subject to the availability of exotic actinide samples.

\subsection{Laser spectroscopy of lawrencium}

The RADRIS technique was successfully applied to nobelium and allowed the identification and the detailed characterization of atomic transitions as discussed in section \ref{sec:nobelium}. 
The technique is suitable for the first laser spectroscopy on the next heavier element lawrencium (Lr, $Z=103$). The cross sections for the production of the Lr isotopes are shown in table~\ref{tab:prod-Lr}. They are in the range of 40--200 nanobarn and thus similar to the cross section for the production of $^{255}$No, which was the lowest among the nobelium isotopes studied so far.

Lr is of special interest as the configuration of the atomic ground state is predicted to be 7s$^2$\,7p~$^2$P$^{\circ } _{1/2}$, unlike its isoelectronic homologue element Lu with a configuration 6s$^2$\,5d~$^2$D$_{3/2}$ \cite{Zou2002,Fritzsche2007,Dzuba2014,Borschevsky2007a}. In this respect, laser spectroscopy may provide experimental confirmation and a theory-independent assessment of this key element property. For the Lr atom there are several predictions for the location of atomic levels that are within reach of laser spectroscopy. The two strongest transitions are predicted to be located in the visible spectral range near $20\,400\,$cm$^{-1}$ (excitation to the 7s$^2$8s state) and in the UV range at about $28\,100\,$cm$^{-1}$ (excitation to the 7s$^2$ 7d state).

\begin{table}[htp]
\caption{Production of Lr isotopes of interest}
\label{tab:prod-Lr}
\begin{center}
\begin{tabular}{|c|c|c|c|}
\hline
Nuclide & Half-life / s & Reaction & Cross Section / nb \\
\hline
$^{254}$Lr & 18.4 & $^{48}$Ca($^{209}$Bi, 3n) & 40 \\
$^{255}$Lr & 31 & $^{48}$Ca($^{209}$Bi, 2n) & 200 \\
$^{256}$Lr & 27 & $^{48}$Ca($^{209}$Bi, 1n) & 60 \\
\hline
\end{tabular}
\end{center}
\label{tab:Lr}
\end{table}

Nevertheless, there are some additional experimental challenges arising for this element. These are summarized in the following.
Lr has a closed f-shell with the valence electrons accessing the p- and d- shells. This results in a reduced ionization potential. In fact the Lr ionization potential is the lowest among all actinides with a value of $4.96\,$eV \cite{Sato2015,Sato2018}. 
Due to an increased desorption enthalpy, which was recently theoretically predicted \cite{Pershina2020}, an increased temperature for the desorption from the Ta filament is required. In combination with the reduced ionization potential this results in an increased background from surface ionization compared to the situation in nobelium. Recently, a detailed study of the filament desorption with stable Yb and Lu was carried out \cite{Murboeck2020} indicating that Hf is a suitable filament material for Lr spectroscopy. In the near future the element lawrencium will be thus tackled with the RADRIS method using Hf filaments. The increasing refractory character of the elements above Lr limits the applicability of the RADRIS approach.  

An alternative approach is the employment of the in-gas jet technique as discussed for the low-energy front-end for the in-flight-separator S$^3$ \cite{Ferrer2013,Ac_ferrer2017towards}.
Here the recoil ions are stopped and neutralized in argon gas and transported by a gas flow.
Resonance ionization spectroscopy can take place either inside the gas cell or in the supersonic gas jet emerging from the gas cell, while the resulting laser ions are guided by a radio frequency quadrupole to a detection station.

\subsection{Ion mobility-based experiments}
\label{sec:mobility}

As mentioned in Sec.~\ref{sec:buffer_gas}, many investigations of the heaviest elements necessitate the production of low-energy radioactive beams and the usage of in-flight separators for element isolation~\cite{Tuerler2015,Block2015,Backe2015}.
Two prominent examples are the mass measurements on nobelium and lawrencium~\cite{Minaya2012} and the aforementioned laser spectroscopy of nobelium~\cite{Laatiaoui2016}.

Such experiments rely to an ever-increasing extent on using buffer-gas catchers for a quick stopping and manipulation of the fusion products.
With the development of new generation stopping cells, new opportunities opened up for studying electronic configurations by Ion Mobility Spectrometry (IMS) at no additional cost of experimental complexity.

In IMS, the speed of gas-phase ions drifting in a uniform electric field through a buffer gas is measured~\cite{Viehland:2018}. The speed is proportional to the ion mobility, which has been found to be sensitive to changes in the electronic configuration, like when an additional orbital is occupied.
This has motivated systematic IMS measurements across the lanthanides~\cite{Laatiaoui2012,Manard:2017} where deviations in electronic configurations were expected to exist for singly-charged ions caused by relativistic effects~\cite{Indelicato2007}. The measurements have proven the ion mobility to be sensitive to the occupation of the 5d orbital while the 4f shell is gradually getting filled~\cite{Laatiaoui:2009}.

Currently, there is ongoing work to extend these IMS measurements to the actinides~\cite{Rickert2020} to deepen the understanding of ion mobility changes in the actinide region.

An added motivation is given by the dependence of the ion mobility on the buffer-gas temperature $T$ and the $E/n_0$ ratio of electric field strength to gas number density. This dependency provides a detailed "image" of the underlying ion-atom interaction potential~\cite{Mason:1988}. Within the different $E/n_0$ ranges, the ion mobility is sensitive to the position and slope of the repulsive wall, the equilibrium distance and the attractive long-range interactions, which can be inferred from modern \emph{ab-initio} electronic structure theory.

Lacking experimental data, however, theoretical approaches are not yet capable of addressing the transport properties of the heaviest elements in a comprehensive and straightforward way.
Only in the last few years theory started to provide quantitative results for weak interatomic interactions involving some lanthanide~\cite{Dalgarno2007,Buchachenko2014,Buchachenko2019} and trans-lanthanide~\cite{Viehland2006,Visentin2020} atoms and ions. While the recent systematic measurements across the lanthanides at room temperature have indicated reasonable accuracy of the scalar-relativistic \emph{ab-initio} interaction potentials~\cite{Viehland:2018}, only additional data covering a wide parameter space of $T$ and $E/n_0$ can fully disclose to what extent a precise description of induction and dispersion forces in the monomers becomes important~\cite{Visentin2020,Rickert2020}.

Hence, systematic IMS studies across the actinides may provide benchmark data for state-of-the-art \emph{ab-initio} calculations and bear a great potential for element identification techniques at minute production quantities. Once experimentally established, trace detection and isotope identification capability of IMS may be of interest for actinide detection in environmental and security applications as well.

Another important application has recently been proposed in which spectroscopic transitions are indirectly detected by using the field-induced drift to discriminate ions in ground and excited electronic states~\cite{Laatiaoui:2020,Laatiaoui2020a}. Even though this electronic-state chromatography effect is well established in the lighter transition metals~\cite{Kemper:1991,Taylor:1999,Iceman:2007}, its investigation in the region of the superheavy elements may provide an alternate way of optical spectroscopy aside of resonance ionization and fluorescence-based techniques.

\subsection{Laser spectroscopy in ion traps}

Laser spectroscopy in ion traps has been performed for many decades and is also discussed in some textbooks, see for example \cite{Major2006,Werth2009}. The confinement and manipulation of even single particles with the help of electromagnetic fields is a backbone of many ion trap-based experiments. In an ion trap a charged particle is confined by electric and magnetic fields in different configurations, most of the time in ultrahigh vacuum. Already in the early days the interaction of trapped ions with laser light was utilized to cool the ions to very low temperature and to perform precision laser spectroscopy. Already in the 1980s the first optical detection of a single ion was achieved \cite{Neuhauser1980}. One of the limitations in optical spectroscopy, the Doppler broadening, can be overcome in ion traps for example by laser cooling to reduce the ion's kinetic energy and the velocity spread. Under these conditions, one can confine trapped ions in a small volume in space that facilitates a detection of photons with the help of imaging charge-coupled device (CCD) cameras.  Thus, ion traps feature the ultimate sensitivity to work with single ions, but their application has for a long time been restricted to off-line measurements of stable and long-lived nuclides. 

A critical step that prevented on-line experiments with radionuclides was the efficient injection of the radioactive ions of interest into the trap. Similar to the discussion in section \ref{sec:buffer_gas} the solution is related to buffer-gas techniques. Besides the gas stopping cells, the introduction of radiofrequency beam cooler and buncher devices, see for example the pioneering work at ISOLTRAP \cite{Herfurth2003}, has opened up new possibilities. While this approach was initially introduced for precision mass measurements, new options also for laser spectroscopy emerge. At JYFL bunched radioactive ions were utilized to improve the sensitivity of collinear laser spectroscopy \cite{Nieminen2002} and, furthermore, optical manipulation within the cooler has been utilized \cite{Cheal2009}.

A related recent development aiming at laser spectroscopy using an ion manipulation device is the MIRACLS project \cite{Sels2020}. In this method a multi-reflection time-of-flight mass spectrometer (MR-ToF-MS) is used for a variant of collinear laser spectroscopy. In addition to the collinear laser spectroscopy using RF cooler bunchers, MR-ToF-MS does not exploit only ion bunching to reduce background from scattered laser photons, but also the feature that an ion bunch can be reflected back and forth between the electrostatic mirrors of the device to increase sensitivity.

Preparing actinide ions for trapping with the help of buffer-gas stopping cells, radio-frequency cooler bunchers and other tools has been achieved, for example in the case of SHIPTRAP \cite{Block2015}. A slowing down and cooling of rare isotopes can be accomplished on a time scale of about 100 milliseconds. Thus, for many of the actinide and transactinide isotopes known to date it will be feasible to prepare them and transfer them to a spectroscopy trap, for example a linear Paul trap.

Thus, in the future ion traps and advanced ion manipulation techniques may be more and more exploited for laser spectroscopy experiments on actinides. This approach may be most suitable for cases where precision measurements of specific observables are desired and atomic transitions have been identified before. In addition, systems with atomic transitions in accessible wavelength ranges will be needed. An attractive option may be the application of laser microwave double resonance techniques to perform hyperfine spectroscopy. In this case the laser excitation is used to pump the population into a specific substate from which a microwave excitation can be performed. In the so-called Lamb-Dicke regime the first-order Doppler effect vanishes providing high resolution. This technique has been used for example for g-factor measurements in barium and europium isotopes \cite{Marx1998,Trapp2000}.

\section{Conclusions}

In recent years, significant advances in the studies of actinide elements have been made. Based on the development of novel and the improvements of established laser spectroscopy methods many new measurements of atomic and nuclear properties of actinide isotopes up to nobelium have been achieved. This trend will continue exploiting novel approaches, new installations and upcoming next-generation RIB facilities. In parallel progress in theory, both in atomic and nuclear physics, has been made benefitting from increased computational capabilities and the development of new methods. This has helped the interpretation of the obtained results and supported the planning of future experiments. 
Despite all these achievements the progress towards higher $Z$ will take time due to more complex atomic configurations and the ever-decreasing yield.

\section*{Acknowledgements}

M. L. acknowledges funding from the European Research Council (ERC) under the European Union’s Horizon 2020 research and innovation program (grant agreement No. 819957). This work was funded in part by the German Federal Ministry of Education and Research (Bundesministerium für Bildung und Forschung BMBF) under grant 05P18UMFN1.


\end{document}